\documentclass{lmcs}
\pdfoutput=1
\usepackage[utf8]{inputenc}

\usepackage{lastpage}
\lmcsdoi{21}{3}{19}
\lmcsheading{}{\pageref{LastPage}}{}{}%
{Jan.~24,~2019}{Sep.~03,~2025}{}

\title{Totality for Mixed Inductive and Coinductive types}

\author[P.~Hyvernat]{Pierre Hyvernat}
\address{
Universit\'e Savoie Mont Blanc, CNRS, LAMA, 73000
Chamb\'ery, France.
}
\email{\href{mailto:pierre.hyvernat@univ-smb.fr}{pierre.hyvernat@univ-smb.fr}}
\urladdr{\url{http://lama.univ-savoie.fr/~hyvernat/}}

\thanks{This work was partially funded by ANR project RECIPROG, project
reference ANR-21-CE48-019-01}


\usepackage{alltt}

\usepackage{stmaryrd}
\usepackage{amssymb,amsmath}
\usepackage{cmll}
\usepackage{xspace}

\newif\ifUSEEXTERNALGRAPHICS
\USEEXTERNALGRAPHICStrue
\ifUSEEXTERNALGRAPHICS
  \usepackage{tikzexternal}
  \tikzsetexternalprefix{figures/}

\else
  \usepackage{tikz}
  \usetikzlibrary{matrix,arrows,calc,shapes.geometric,positioning,decorations.pathmorphing,decorations.pathreplacing}

  \usetikzlibrary{external}
  \tikzsetexternalprefix{figures/}
\fi

\let\nodeSize=\scriptsize
\let\transitionSize=\scriptsize
\newcommand\Set{\mathsf{Set}}

\newcommand\Tot{\mathsf{Tot}}
\newcommand\seq[1]{\ensuremath{\overline{#1\,}}}

\newcommand\sub{\subseteq}

\newcommand\V{\ensuremath{\mathcal{V}}\xspace}

\newcommand\StdSem{\Theta^{\text{\rm std}}}

\newcommand\Sem[1]{\left\llbracket#1\right\rrbracket}

\newcommand\BLANK{{\texttt{\char"5F}}}

\newcommand\iso{\cong}

\renewcommand\tt[1]{\leavevmode\text{\textup{\texttt{#1}}}}

\newcommand\x{{\tt{x}}\xspace}
\newcommand\y{{\tt{y}}\xspace}

\newcommand\f{{\tt{f}}\xspace}
\newcommand\g{{\tt{g}}\xspace}
\newcommand\C{{\tt{C}}\xspace}
\newcommand\DD[1]{{\tt{.#1}}\xspace}
\newcommand\D{{\DD{D}}\xspace}

\newcommand\p[1]{{\ensuremath{^{#1}}}}

\newcommand\T[1]{{\left|#1\right|}}
\newcommand\Win{\mathcal{W}}
\newcommand\chariot{\texttt{chariot}\xspace}
\newcommand\bstruct{\tt{\symbol{`\{}}}
\newcommand\estruct{\tt{\symbol{`\}}}}
\newcommand\st[1]{%
 \bgroup%
  \renewcommand\DD[1]{{\tt{##1}}\xspace}%
   \renewcommand\D{{\DD{D}}}%
    \bstruct{}#1\estruct%
     \egroup}
\newcommand\transition[1]{\xrightarrow{#1}}

\makeatletter
\newlength{\negph@wd}
\DeclareRobustCommand{\negphantom}[1]{%
  \ifmmode
    \mathpalette\negph@math{#1}%
  \else
    \negph@do{#1}%
  \fi
}
\newcommand{\negph@math}[2]{\negph@do{$\m@th#1#2$}}
\newcommand{\negph@do}[1]{%
  \settowidth{\negph@wd}{#1}%
  \hspace*{-\negph@wd}%
}

\DeclareRobustCommand{\dirsup}{%
  \@ifnextchar_{\sub@bigsqcupwithup}{\nosub@bigsqcupwithup}%
}

\newcommand{\nosub@bigsqcupwithup}{%
  \mathop{
    \mathchoice
      {\disp@bigsqcupwithup{}}
      {\nodisp@bigsqcupwithup{\hphantom{\scriptscriptstyle\uparrow}}}
      {\nodisp@bigsqcupwithup{\hphantom{\scriptscriptstyle\uparrow}}}
      {\nodisp@bigsqcupwithup{\hphantom{\scriptscriptstyle\uparrow}}}
  }
}
\def\sub@bigsqcupwithup_#1{%
  \mathop{
    \mathchoice
      {\disp@bigsqcupwithup{#1}}
      {\nodisp@bigsqcupwithup{#1}}
      {\nodisp@bigsqcupwithup{#1}}
      {\nodisp@bigsqcupwithup{#1}}
  }
}

\newcommand{\disp@bigsqcupwithup}[1]{%
  \negphantom{{}^\uparrow}%
  {\mathop{\hphantom{{}^\uparrow}{\bigsqcup}{}^\uparrow}\limits_{#1}}%
}
\newcommand{\nodisp@bigsqcupwithup}[1]{%
  \bigsqcup^{\scriptscriptstyle\uparrow}_{#1}%
}
\makeatother

\newcommand\Rule[2]{\frac{\phantom{\big(}\quad{\displaystyle #1}\quad}{\phantom{\Big(}\quad{\displaystyle #2}\quad}}

\newenvironment{myequation}{%
  \ignorespaces\[\everymath{\displaystyle}\begin{array}{rclr}%
    }{%
  \end{array}\]\ignorespacesafterend%
}

\newcommand\ie{i.e.\ }

\newcommand\cf{c.f.\ }

\begin{document}

\begin{abstract} %%%<<<1
  This paper introduces an ML / Haskell like programming language with nested
  inductive and coinductive algebraic datatypes called \chariot. Functions are
  defined by arbitrary recursive definitions and can thus lead to
  non-termination and other ``bad'' behavior. \chariot comes with a
  \emph{totality checker} that tags possibly ill-behaved definitions. Such a totality
  checker is mandatory in the context of proof assistants based on type theory
  like Agda.
  Proving correctness of this checker is far from trivial and relies on
  \begin{enumerate}
    \item an interpretation of types as parity games,
    \item an interpretation of correct values as winning strategies for those games,
    \item the Lee, Jones and Ben Amram's size-change principle, used to check
      that the strategies induced by recursive definitions are winning.
  \end{enumerate}
  % Using parity games is important as it is known that when inductive and
  % coinductive types are nested, termination and productivity aren't enough to
  % ensure soundness.
  This paper develops the first two points, the last step being the subject of
  an upcoming paper.
  A prototype has been implemented and can be used to experiment with the
  resulting totality checker, giving a practical argument in favor of this
  principle.

\end{abstract} %%%>>>1

\maketitle

\begingroup\footnotesize
\makeatletter
\def\l@section{\@tocline{2}{0pt}{1pc}{5pc}{}}
\def\l@subsection{\@tocline{2}{0pt}{2pc}{6pc}{}}
\makeatother
% \tableofcontents
\endgroup

\section*{Introduction}

%%% intro <<<1
Inductive types (also called algebraic datatypes) are a cornerstone of typed
functional programming: Haskell and Caml both rely heavily on them. One
mismatch between the two languages is that Haskell is \emph{lazy} while Caml
is \emph{strict}. A definition like
{\small\begin{alltt}
    let rec nats : nat -> nat list
                 = fun n -> n::(nats (n+1))
\end{alltt}}
\noindent
is valid but useless in Caml because the evaluation mechanism will loop trying
to evaluate it completely (call-by-value evaluation), resulting in a stack
overflow exception. In Haskell, because evaluation is lazy (call-by-need),
such a definition isn't unfolded until strictly necessary and asking for its
third element will only unfold the definition three times.
Naively, it seems that types in Caml correspond to ``least fixed points''
while they correspond to ``greatest fixed points'' in Haskell.
% The difference is subtler than that: if one interprets types as domains of
% values, then the distinction doesn't exist: domain equations have unique
% solutions.

The aim of this paper is to introduce a language, called
\chariot,\footnote{All the examples will now be given using the syntax of
\chariot which is described in sections~\ref{sub:types} and~\ref{sub:rec_def}.
They should be readable by anyone with a modicum of experience in
functional programming. A prototype implementation in Caml is available from
\url{https://github.com/phyver/chariot} for anyone wishing to experiment with
it.} which distinguishes between least and greatest fixed points and where
the user can nest them arbitrarily to define new datatypes. To offer a
familiar programming experience, definitions are not restricted and any
well-typed recursive definition is allowed. In particular, it is possible to
write badly behaved definitions like
{\small\begin{alltt}
    val f : nat -> nat
      | f 0 = 1
      | f (n+1) = f(f n)         \textit{-- f(1) => f(f(0)) => f(1) => ...}
\end{alltt}}

\noindent
To guarantee that a definition is correct, two independent steps are
necessary:
\begin{enumerate}
  \item Hindley-Milner type-checking~\cite{milner:polymorphism} to guarantee that
    evaluation doesn't provoke runtime errors,
  \item a \emph{totality test} to check that the definition is valid with
    respect to the coinductive / inductive types it involves.
\end{enumerate}
In simple cases that only involve inductive types, a typed value is total when
it is finite, and a function is total if it sends finite values to finite
values, \ie it terminates.\footnote{The case of general W-types is slightly
more complex: even if they can contain infinite values, those are necessarily
well-founded. The version of \chariot presented here doesn't allow such types
but extending the theory is certainly possible. A simple example is given on
page~\pageref{sub:conclusion_HOT} in the conclusion.} In the presence of
coinductive types however, values can be infinite. Totality then roughly means
that the only infinite parts of a value come from coinduction. As the examples
will show, this quickly becomes very subtle when inductive and coinductive
types are interleaved.

The totality test generalizes an older termination checker~\cite{ph:SCT}.
Like before, any definition that passes this test is guaranteed to be correct
but because the halting problem is undecidable, some correct definitions are
rejected. In a programming context, the programmer may choose to ignore the
warning if she (thinks she) knows better. In a proof-assistant context
however, it cannot be ignored as non-total definitions lead to
inconsistencies, the most obvious example being
{\small\begin{alltt}
    val magic_proof = magic_proof
\end{alltt}}
\noindent
which is non-terminating but belongs to all types. As we'll see, there are
subtler examples of definitions that normalize to values but still lead to
inconsistencies.

In Coq~\cite{coq_manual}, the productivity condition for coinductive
definitions is ensured by a strict syntactic condition
(guardedness~\cite{coquand:infinite}) similar to the condition that inductive
definitions need to have one structurally decreasing argument.
In Agda~\cite{agda_manual}, the user can write arbitrary recursive definitions
and the productivity condition is ensured by the termination checker. Agda's
checker extends the termination checker developed by A. Abel~\cite{abel:foetus}
to deal with coinductive types, but while this is sound for simple types like
streams, it is known to be unsound for nested coinductive and inductive
types~\cite[Section~5]{altenkirch_danielsson:nested}. Agda's totality checker
is patched to deal with such counter examples but as far as I know, no proof of
correctness is available.\footnote{Their counterexample is described in
Section~\ref{sec:CounterExample} on page~\pageref{sec:CounterExample}. Simply
put, the problem of checking termination and productivity independently is that
we cannot distinguish between~$\mu_X\nu_Y.F(X,Y)$ and~$\nu_Y\mu_X.F(X,Y)$,
which have different semantics.}
This paper provides a combinatorial characterization of totality. An upcoming
paper will describe how it can be used to implement a totality checker.
%%%>>>1

\subsection*{Related Works} %%%<<<1

\subsubsection*{Circular proofs} %%%<<<2

The main inspiration for this work comes from ideas developed by
L.~Santocanale in his work on circular
proofs~\cite{santocanale:mu_bicomplete,santocanale:circular_proofs,santocanale:parity}. Circular
proofs are defined for a linear proof system and are interpreted in categories
with products, coproducts and enough initial algebras / terminal coalgebras.
In fine, the criterion implemented in \chariot uses a strong combinatorial
principle (the size-change principle) to check a sanity condition on a kind of
circular proof (a program). This is strictly stronger than the initial
criterion used by L. Santocanale and G. Fortier, which corresponds to the
syntactical structurally decreasing / guardedness condition on recursive
definitions.

However, while circular proofs were a primary inspiration, the~\chariot
language cannot be reduced to a circular proof system. The main problem is
that existing circular proof systems are linear and do not have a simple
cut-elimination procedure, \ie an evaluation mechanism. Cuts and exponentials
would be needed to interpret the full \chariot language and while cuts can be
added~\cite{fortier_santocanale:cuts,fortier:PHD}, adding exponentials looks difficult and
hasn't been done.

More recent works in circular proof theory replace L.~Santocanale's criterion
by a much stronger combinatorial
condition~\cite{doumane:PHD,doumane:completeness}. It involves checking that
some infinite words are recognized by a parity automata, which is a decidable
problem. The presence of parity automata points to a relation between this
work and the present paper, but the different contexts make it all but
obvious.

%%%>>>2

\subsubsection*{Size-change principle} %%%<<<2

The second idea will be developed in an upcoming paper and consists of
adapting the \emph{size-change principle} (SCP) from C. S. Lee, N. D. Jones
and A. M. Ben-Amram~\cite{lee_jones_benamram:SCT} to the task of checking
general totality. This problem is subtle as totality is strictly more than
termination and productivity.\footnote{A nice counter example, due to N. A.
Danielsson and T. Altenkirch~\cite{altenkirch_danielsson:nested} is given on
page~\pageref{sec:CounterExample}.} Moreover, while the principle used to
check termination of ML-like recursive definitions~\cite{ph:SCT} was
inherently untyped, totality checking needs to be somewhat type aware. For
example, in \chariot, records are lazy and are used to define coinductive
types. The definition
{\small\begin{alltt}
      val inf = Node \st{ Left = inf; Right = inf }         \textit{-- infinite binary tree}
\end{alltt}}
\noindent
yields an infinite binary tree and depending on the types of~$\tt{Node}$,
${\tt{Left}}$ and~$\tt{Right}$, the definition may be correct or incorrect (\cf
page~\pageref{ex:inf})!

% This work extends an earlier termination criterion~\cite{ph:SCT} to check
% totality. Some minor differences are
% \begin{itemize}
%   \item
%     anonymous tuples are replaced with records with named fields,

%   \item
%     the notion of weight was simpler as priorities were not necessary,

%   \item
%     the approximation order has been reversed to coincide with the domain
%     order of the last section,

%   \item
%     sums are now written with a product notation in order to keep the symbol
%     ``$+$'' for non-deterministic sums.
% \end{itemize}
%%%>>>2

\subsubsection*{Charity} %%%<<<2

The closest ancestor to \chariot is the language \tt{charity}\footnote{By the
way, the name \chariot was chosen as a reminder of this
genealogy.}~\cite{cockett:charity,cockett:charitable}, developed by R. Cockett and T.
Fuku\-shima. It lets the programmer define types with arbitrary nesting
of induction and co\-induction. Values in these types are defined using
categorical principles.

\begin{itemize}
  \item Inductive types are \emph{initial} algebras: defining a function
    \emph{from} an inductive type amounts to defining an algebra for the
    corresponding operator.

  \item Coinductive types are \emph{terminal} coalgebras: defining a
    function \emph{to} an inductive type amount to defining a coalgebra for
    the corresponding operator.
\end{itemize}
It means that recursive functions can only be defined via eliminators. By
construction, they are either ``trivially'' structurally decreasing on their
argument, or ``trivially'' guarded. The advantage is that {all} functions
are total by construction and the disadvantage is that the language is not
Turing complete.
%%%>>>2

\subsubsection*{Guarded recursion} %%%<<<2

Another approach to checking correctness of recursive definitions is based on
``guarded recursion'', initiated by H. Nakano~\cite{nakano:guarded} and later extended
in several directions~\cite{birkedal:guarded,guatto:guarded}. In this approach, a new modality
``later'', written~``$\triangleright$'', is introduced. The
type~``$\triangleright T$'' gives a syntactical way to talk about terms that
``will later, after some computation, have type~$T$''. This work is quite
successful and has been extended to very expressive type systems. The drawbacks
are that this requires a non-standard type theory with a not quite standard
denotational semantics (topos of trees). Moreover, it makes programming more
difficult as it introduces new constructors for types and terms. Finally, these
works only consider greatest fixed points (as in Haskell) and are thus of
limited interest for systems like Agda or Coq.

%%% >>>2

\subsubsection*{Sized types} %%%<<<2

This approach extends type theory with a notion of ``size'' that annotate
types. It has been successful and is implemented in
Agda~\cite{abel:miniagda,abel:termination}. It is possible to specify that the \tt{map}
function on list has type~$\forall n,\tt{list}^n(T) \to \tt{list}^n(T)$,
where~$\tt{list}^n(T)$ is the type of lists with~$n$ elements of type~$T$.
These extra parameters give information about recursive functions and make it
easier to check termination. A drawback is that functions on sized-types must
take extra size parameters. This complexity is balanced by the fact that most
of them can be inferred automatically and are thus mostly the libraries'
implementers' job: in many cases, sizes are invisible to the casual user. Note
however that sizes only help showing termination and productivity. Developing
a totality checker is orthogonal to designing an appropriate notion of size
and the totality checker described in this paper can probably work hand in
hand with standard size notions.

It should also be noted that the interaction between sizes and Agda's complex
types is unclear and that, like for the actual termination checker, no
corresponding proof of correctness has ever been written.
%%%>>>2

\subsubsection*{Fixed points in game semantics} %%%<<<2

An important tool in this paper is the notion of \emph{parity game}. P.
Clairambault~\cite{clairambault:strong_functors} explored a category of games enriched with
winning conditions for infinite plays. The way the winning condition is
defined for least and greatest fixed points is reminiscent of L. Santocanale's
work on circular proofs and the corresponding category is cartesian closed.
Because this work is done in a more complex setting and aims for generality,
it seems difficult to extract a practical test for totality from it. The
present paper aims for specificity and practicality by devising a totality
test for the usual semantics of recursion.
%%%>>>2

\subsubsection*{SubML} %%%<<<2

C. Raffalli and R. Lepigre used the size-change principle to check correctness
of recursive definitions in the language SubML~\cite{lepigre_raffalli:subml}. Their approach
uses a powerful but non-standard type theory with many features: subtyping,
polymorphism, sized-types, control operators, some kind of dependent types,
etc.
%All those features were chosen to facilitate the programmer' experience.
On the downside, it makes their type theory more difficult to compare with
other approaches. Note that like in Agda or \chariot, they do allow arbitrary
definitions that are checked by an incomplete totality checker.
% The similarity of the approach isn't surprising considering previous
% collaborations between the authors.
%
One interesting point of their work is that the size-change termination is
only used to check that some object (a proof tree) is well-founded: even
coinductive types are justified with well-founded proofs.
%%%>>>2

%\subsubsection*{Nax} %%%<<<2

%Another programming language with nested inductive / coinductive types is the
%Nax language~\cite{nax:PHD}, based on so called ``Mendler style
%recursion''~\cite{mendler:inductive}. One key difference is that the Nax language is
%very permissive on the definitions of types (it is for example possible to
%define fixed points for non positive type operators) and rather restrictive on
%the definition of values: they are defined using various combinators similar
%(but stronger than) to the way values are defined in \tt{charity}, and usual
%recursive definitions are not allowed. Since no implementation of Nax is
%available, it is however difficult to experiment with it.
%%%%>>>2

%%%>>>1

\subsection*{Plan of the Paper} %%%<<<1

We start by introducing the language \chariot and its denotational semantics
in Section~\ref{section:semantics}. We assume the reader is familiar with
functional programming, recursive definitions and their semantics,
Hindley-Milner type checking, algebraic datatypes, pattern matching, etc. The
notion of totality is also given there. We then describe, in
Section~\ref{section:totality}, a combinatorial approach to totality that
comes from L. Santocanale's work on circular proofs. This reduces checking
totality of a definition to checking that the definitions gives a winning
strategy in a parity game associated to the type of the definition.

The actual totality checker using the size-change principle will be described
in an upcoming paper~\cite{ph:SCP}.
% The first step will consist of developing an abstract interpretation of
% recursive definitions that can accommodate the size-change principle. This
% semantics will be both untyped and non-deterministic. The notion of
% \emph{call-graph}, central to the implementation of the size-change
% principle can be defined on top of that. Applying and implementing the
% size-change principle follows naturally from there.
%%%>>>1

%% vim600:set foldmarker=<<<,>>> foldmethod=marker fileencoding=ascii spelllang=en spell: %%

\section{The Language and its Semantics}
\label{section:semantics}

\subsection{Type Definitions} \label{sub:types}%%%<<<1

Datatypes in \chariot come in two flavors: those corresponding to sum types (\ie
colimits) and those corresponding to product types (\ie limits).
%
% The syntax is itself similar to that of \tt{charity}:
\begin{itemize}
  \item
    an inductive type is specified by a list of \emph{constructors} whose
    \emph{codomain} is the type being defined,

  \item
    a coinductive type is specified by a list of \emph{destructors} whose
    \emph{domain} is the type being defined.
\end{itemize}
They are respectively introduced by the keywords ``\tt{data}'' and
``\tt{codata}'' and may have parameters. Type parameters are written with a
quote as in Caml. Here are some examples:
{\small\begin{alltt}\label{ex:type_defs}
    codata unit where                             --{\it unit type: no destructor, see below}

    codata prod('x,'y) where  Fst : prod('x,'y) -> 'x                     --{\it pairs}
                            | Snd : prod('x,'y) -> 'y

    data nat where  Zero : unit -> nat                     --{\it unary natural numbers}
                  | Succ : nat  -> nat

    data list('x) where  Nil  : unit               -> list('x)        --{\it finite lists}
                       | Cons : prod('x, list('x)) -> list('x)

    codata stream('x) where  Head : stream('x) -> 'x             --{\it infinite streams}
                           | Tail : stream('x) -> stream('x)
\end{alltt}\label{type:stream}}\noindent
Elements of a \tt{codata} type are written using a structure notation, by
giving values to all fields in any order, as in~$\st{\tt{Fst} = u ; \tt{Snd} =
v}$ and the \tt{unit} type given above contains a single value~\st{}. The
\tt{data} type without constructor isn't as useful as it doesn't contain any
value!
Examples will sometimes use shortcuts, allowed in the implementation, and
write \tt{Zero} (instead of~\tt{Zero\st{}}) or~\tt{Cons(x,xs)} (instead
of~\tt{Cons\st{Fst=x;Snd=xs}}).

We also introduce a notion of ``rank'', which correspond to the
order in which the type definitions are introduced. Larger ranks may use
smaller ranks in their definition, like \tt{list} requiring \tt{prod}.
\begin{defi}\label{def:type_defs}
  the sets~$\mathcal{D}_r$ of \emph{definitions of rank~$r$}
  and~$\mathcal{T}_r(P_1, \dots, P_n)$ of \emph{type expressions of rank~$r$
  with parameters among~$P_1, \dots, P_n$} are defined by mutual induction:

  \begin{enumerate}
    \item
      For~$r>0$, the set $\mathcal{D}_r$ contains all definitions of the form
      \begin{center}
      \begin{minipage}[t]{.4\textwidth}
      {\small\begin{alltt}
data S('x, 'y, ...) where
    | \(\C\sb1\) : \(T\sb1\) -> S('x, 'y, ...)
    ...
    | \(\C\sb{k}\) : \(T\sb{k}\) -> S('x, 'y, ...)
    \end{alltt}}
    \end{minipage}
    \begin{minipage}[t]{.1\textwidth}
      and
    \end{minipage}
    \begin{minipage}[t]{.4\textwidth}
    {\small\begin{alltt}
codata S('x, 'y, ...) where
    | \(\tt{D}\sb1\) : S('x, 'y, ...) -> \(T\sb1\)
    ...
    | \(\tt{D}\sb{k}\) : S('x, 'y, ...) -> \(T\sb{k}\)
    \end{alltt}}
      \end{minipage}
      \end{center}
      where
      \begin{itemize}
        \item
          \tt{S} is the name of the type being defined,
        \item
          each~$\tt{C}_i$ is the name of a constructor,
        \item
          each~$\tt{D}_i$ is the name of a destructor,
        \item
          each~$T_i$ is an element of~$\mathcal{T}_{r-1}(\tt{S('x, 'y, ...)},\tt{'x}, \tt{'y}, \dots)$.
      \end{itemize}

    \item
      The set~$\mathcal{T}_r({P}_1, \dots, {P}_n)$ where the~${P}_i$ are
      syntactical parameters is defined by the grammar
      \[
      T \in \mathcal{T}_r({P}_1, \dots, {P}_n)
      \qquad::=\qquad
      {P}_i \quad|\quad \tt{T}(T_1, \dots, T_m)
      \]
      where
      \begin{itemize}
        \item
          \tt{T} is the name of a type in some~$\mathcal{D}_{r'}$ of arity~$m$,
          with~$r'\leq r$,

        \item
          each $T_i$ belongs to~$\mathcal{T}_{r}({P}_1,\dots,{P}_n)$.
      \end{itemize}

    \item
      The rank of a definition [resp. type expression] is the smallest~$r$
      such that the definition is in~$\mathcal{D}_r$ [resp. the type
      expression is in some~$\mathcal{T}_r(\dots)$].

    \item
      The head rank of a type expression is the rank of the definition of its
      head type constructor; or~$0$ if the type expression is a parameter.

  \end{enumerate}
\end{defi}
Strictly speaking, each name~\tt{T} is indexed by its definition and each
constructor / destructor name is indexed by its corresponding type name. In
practice, \chariot forbids the reuse of names so that there is no ambiguity
about which definition defines a given type, or about which type corresponds to
a given constructor / destructor.\footnote{Extending the definition to mutual
type definitions is left as an exercise for the diligent reader! It is
straightforward if one keeps in mind that all types in a mutual definition are
\tt{data} or \tt{codata} and that they all have the same parameters~``\tt{('x,
'y, ...)}''.}

Note that to make the theory slightly simpler, constructors always have a
single argument.
% Expressivity doesn't suffer because we can always use a
% tuple~$\st{\tt{Fst}=t_1; \tt{Snd}=t_2}$ as argument.
Of course, the implementation of \chariot allows constructors of arbitrary
arity and the theory can be extended to deal with this.

\medbreak
Destructors act as projections and because of the universal property of
terminal coalgebras, we think about elements of a codatatype as records. This
is reflected in the syntax of terms. For example, the following defines
(recursively) the stream with infinitely many~$0$s. (The syntax for recursive
definitions will be formally given in Definition~\ref{def:definitions}.)
{\small\begin{alltt}\label{ex:zeros}
    val zeros : stream(nat)
      | zeros = \st{ Head = Zero ; Tail = zeros }
\end{alltt}}\noindent
The denotational semantics of a codata is going to be a \emph{coinductive}
type (greatest fixed-point) while the semantics of a data is going to be an
\emph{inductive} type (least fixed-point). In order to have a sound
operational semantics, codata should not be fully evaluated. The easiest way
to ensure that is to stop evaluation on records: evaluating ``\tt{zeros}'' will
result in~``\tt{\st{Head = $\Box$; Tail = $\Box$}}'' where the~``$\Box$'' are non
evaluated chunks.
The copattern view~\cite{abel:copatterns} is natural here. The
definition of~\tt{zeros} using copatterns (allowed in \chariot) looks like
{\small\begin{alltt}
    val zeros : stream(nat)
      | zeros.Head = Zero
      | zeros.Tail = zeros
\end{alltt}}\noindent
We can interpret the clauses as a terminating rewriting system. In particular,
the term \tt{zeros} doesn't reduce by itself.
Because this paper is only interested in the denotational semantics of
definitions, the details of the evaluation mechanism are fortunately
irrelevant. % and the two definitions of~\tt{nats} are equivalent.

\bigbreak
We will use the following conventions:
\begin{itemize}
  \item
    outside of actual type definitions (given using \chariot's syntax), type
    parameters will be written without quote: $\x$, $\x_1$, \dots

  \item
    an unknown datatype will be written~$\tt{T}_{\mu}(\x_1,\dots,\x_k)$ and an
    unknown codatatype will be written~$\tt{T}_{\nu}(\x_1,\dots,\x_k)$,

  \item
    an unknown type of unspecified polarity will be
    written~$\tt{T}(\x_1,\dots,\x_k)$.
\end{itemize}
%%%>>>1

\subsection{Values} %%%<<<1

% Given a recursive definition, we are interested in the ``healthiness'' of its
% semantics. Such considerations take place in the realm of semantics values and
% while every reader will have her favorite programming language and reduction
% strategy, those are mostly irrelevant to the rest of the paper.

Any finite list of recursive definitions only involves a finite number of
types, with a finite number of constructors and destructors. We thus fix, once
and for all, a finite set of constructor and destructor names. Because we
deal with semantically infinite values, the next definition is of course
coinductive.
\begin{defi}\label{def:values}\leavevmode

  \begin{enumerate}
    \item
  The set of \emph{values with leaves in~$X_1$,\dots, $X_n$},
  written~$\V(X_1,\dots, X_n)$ is defined coinductively by the grammar
  \[
    v \qquad::=\qquad
    \bot \quad|\quad
    x \quad|\quad
    \C\,v \quad|\quad
    \st{\D_1=v_1; \dots; \D_k=v_k}
  \]
  where
  \begin{itemize}
    \item
      each $x$ is in one of the~$X_i$,

    \item
      each $\C$ belongs to a fixed finite set of \emph{constructors},

    \item
      each $\tt{D}_i$ belongs to a fixed finite set of \emph{destructors},

    \item
      the order of fields inside records is unimportant,

    \item
      $k$ can be 0.
  \end{itemize}

  \item
  If the~$X_i$ are ordered sets, the order on~$\V(X_1, \dots,X_n)$ is
  generated by
  \begin{itemize}
    \item
      $\bot \leq v$ for all values $v$,

    \item
      if $x\leq x'$ in $X_i$, then $x\leq x'$ in $\V(X_1, \dots, X_n)$,

    % \item\label{item:record_subtyping}
    %   $\st{\D_1=v_1; \dots; \D_k=v_{k+l}} \leq \st{\D_1=v_1; \dots;
    %   kD_k=v_{k}}$,

    \item
      ``$\leq$'' is contextual: if $u\leq v$ then $C[\x:=u] \leq C[\x:=v]$ for any
      value~$C$ containing a formal variable~$\x$ (substitution is defined in
      the obvious way).
  \end{itemize}

  \end{enumerate}
\end{defi}
The set of values (first point in Definition~\ref{def:values}) is defined
coinductively but the order on values (second point in
Definition~\ref{def:values}) is defined inductively. Reasoning about about
the order is usually done using simple inductive proofs.

%%%>>>1

\subsection{Semantics in Domains} %%%<<<1

There is a natural interpretation of types in the category of algebraic DCPOs
where morphisms are continuous functions that are \emph{not} required to
preserve the least element. An algebraic DCPO is an order with the following
properties:

\begin{itemize}
 \item
   every directed set has a least upper bound (DCPO),

 \item
   it has a basis of compact elements (algebraic).

\end{itemize}
Unless specified otherwise, ``domain'' will always refer to an algebraic DCPO.
Recall that any partial order can be completed to a DCPO whose compact
elements are exactly the element of the partial order. This \emph{ideal
completion} formally adds limits of all directed sets.
The following can be proved directly but is also a direct consequence of this
general construction.
\begin{lem}\label{lem:values_domain} If the~$X_i$s are domains, then
  $\big(\V(X_1, \dots, X_n),\leq\big)$ is a domain.
\end{lem}

Given a type definition of arity~$n$ in~$\mathcal{D}_r$, its semantics
sends the sets~$X_1, \dots, X_n$ to a subset of values
in~$\V(X_1, \dots, X_n)$.
\begin{defi}\label{def:semantics_type}
  We define, by mutual induction, the interpretation of type definitions and
  type expressions:

  \begin{itemize}
    \item 
      Let~$\tt T\in\mathcal{D}_r$ be a type definition of arity~$n$.
      The interpretation of~\tt{T}, written~$\Sem{\tt T}$, sends any sequence~$\seq
      X=X_1, \dots, X_n$ of~$n$ sets to a subset of~$\V(\seq X)$.
      It is defined \emph{coinductively} by the following ``typing'' rules:
      \begin{enumerate}
        \item
          $\displaystyle\Rule{}{\bot:\Sem{\tt T}(\seq{X})}{}$,

        \item
          $\displaystyle\Rule{u\in X_i}{u:\Sem{\tt T}(\seq X)}{}$,

        \item
          $\displaystyle\Rule{u:\Sem{S}\big(\Sem{\tt{T}_\mu}(\seq X), \seq X\big)}{\C\,u : \Sem{\tt{T}_{\mu}}(\seq X)}{}$
          when~$\C:S \to \tt{T}_{\mu}(\seq\x)$ is a constructor
          of~$\tt{T}_{\mu}$,\footnote{There and in point~(4), the
          interpretation~$\Sem{T}$ is already defined as its
          rank is strictly smaller than~$r$.}

        \item
          $\displaystyle\Rule{u_1:\Sem{S_1}\big(\Sem{\tt{T}_\nu}(\seq X),\seq X\big) \quad \dots \quad u_k:\Sem{S_k}\big(\Sem{\tt{T}_\nu}(\seq X),\seq X\big)}
          {\st{\D_1 = u_1;\ \dots;\ \D_k = u_k}: \Sem{\tt{T}_\nu}(\seq X)}{}$
          if the destructors of~$\tt{T}_{\nu}$ are exactly
          the~$\tt{D}_i:\tt{T}_{\nu}(\seq\x) \to S_i$, for~$1\leq i\leq k$.
      \end{enumerate}

    \item
      the interpretation of a type
      expression~$T\in\mathcal{T}_r(\seq\x)$ is defined by induction:
      \begin{itemize}
        \item
          if~$T=\x_i$, then~$\Sem{T}(\seq X) = X_i$,

        \item
          if~$T=\tt{T}(S_1, \dots, S_m)$ then~$\Sem{T}(\seq X) = \Sem{\tt T}\big(\Sem{S_1}(\seq X), \dots,\Sem{S_m}(\seq X)\big)$.
      \end{itemize}
  \end{itemize}
\end{defi}
% Equivalently,~$\Sem{T}(\seq X)$ could be defined as the ideal completion of its
% compact elements, obtained \emph{inductively} when the second rule is
% restricted to compact elements of the parameters.
%
\noindent
The following is easily proved by induction on~$T$.
\begin{prop}\label{prop:semantics_fixed_points}
  Let~$\seq X=X_1$,\dots, $X_n$ be domains, if~$\tt T$ is a type then
  \begin{enumerate}
    \item
      with the order inherited from~$\seq X$ (point~(2) in
      Definition~\ref{def:values}), $\Sem{\tt T}(\seq X)$ is a domain,

    \item
      $\seq X \mapsto \Sem{\tt T}(\seq X)$ is functorial.

    \item
      for a datatype $\tt{T}_{\mu}(\seq\x)$ with
      constructors~$\C_i:S_i \to \tt T_\mu(\seq\x)$, we have
      \begin{myequation}
         \Sem{\tt T_\mu}\big(\seq X\big)
         & = &
           \{\bot\} \ \cup
         \bigcup_{i=1,\dots, k}
         \Big\{
           \C_i u_i \mid
           u_i\in\Sem{S_i}\big(\Sem{\tt{T}_\mu}(\seq X),\seq X\big)
         \Big\}
         \\
         &\iso&
         \Big(\Sem{S_1}\big(\Sem{\tt{T}_\mu}(\seq X),\seq X\big) + \cdots + \Sem{S_k}\big(\Sem{\tt{T}_\mu}(\seq X),\seq X\big)\Big)_\bot\\
      \end{myequation}

    \item
      for a codatatype $\tt{T}_{\nu}(\seq\x)$ with
      destructors~$\tt{D}_i:S_i \to \tt T_\nu(\seq\x)$, we have
      \begin{myequation}
         \Sem{\tt T_\nu}\big(\seq X\big)
         & = &
           \{\bot\} \cup
         \Big\{
           \st{\dots; \D_i=u_i; \dots} \mid \text{$i=1,\dots,k$ and  $u_i\in\Sem{S_i}\big(\Sem{\tt{T}_\nu}(\seq X),\seq X\big)$}
         \Big\}
         \\
         &\iso&
         \Big(\Sem{S_1}\big(\Sem{\tt{T}_\nu}(\seq X),\seq X\big) \times \cdots \times \Sem{S_k}\big(\Sem{\tt{T}_\nu}(\seq X),\seq X\big)\Big)_\bot\\
      \end{myequation}
  \end{enumerate}
\end{prop}\noindent
The operations~$+$ and~$\times$ are the set theoretic operations (disjoint
union and cartesian product), and~$S_\bot$ is the usual notation
for~$S\cup\{\bot\}$. This shows that the
semantics of types are fixed points of standard operators.
For example, $\Sem{\tt{nat}}$ is the domain of ``lazy natural numbers'':
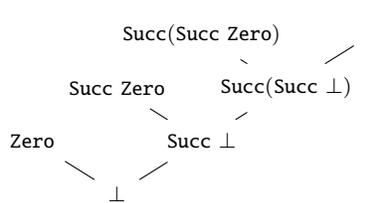
\begin{center}
  \begin{tikzpicture}[baseline=(current bounding box.center),
       level distance=25pt,
       sibling distance=80pt,
       emptynode/.style={circle, inner sep=1.5pt, fill=black},
       treenode/.style={inner sep=5pt},
       font=\scriptsize,
       grow=up,
       % level 1/.style ={level distance=15pt},
       % level 2/.style ={level distance=25pt},
       % level 3/.style ={level distance=30pt},
       scale=.8
     ]
     \coordinate
     node[treenode] {$\bot$}
       child{node[treenode] {$\tt{Succ}\ \bot$}
         child{node[treenode] {$\tt{Succ} (\tt{Succ}\ \bot)$}
           child{node[treenode] {\raise20pt\hbox{$\vdots$}}}
           child {node[treenode] {$\tt{Succ} (\tt{Succ}\  \tt{Zero})$}}
         }
         child {node[treenode] {$\tt{Succ}\ \tt{Zero}$}}
       }
       child {node[treenode] {$\tt{Zero}$}}
       ;
   \end{tikzpicture}
\end{center}
and the following are different elements of $\Sem{\tt{stream(nat)}}$:
\begin{itemize}
  \item
    $\bot$,

  % \item
  %   $\{\tt{Head} = \bot;\tt{Tail} = \bot\}$,

  \item
    $\{\tt{Head} = \tt{Succ}\,\bot; \tt{Tail} = \bot\}$

  \item
    $\{\tt{Head} = \tt{Zero}; \tt{Tail} = \{\tt{Head} = \tt{Zero}; \tt{Tail} =
     \{\tt{Head} = \tt{Zero}; \dots \}\}\}$
\end{itemize}
%%%>>>1

\subsection{Semantics in Domains with Totality} %%%<<<1

We use domains to be able to use Kleene's fixed point theorem to define the
interpretation of recursive definitions (next section). Unfortunately, those
domains cannot distinguish greatest and least fixed point: the functors
defined by types are \emph{algebraically
compact}~\cite{barr:algebraically_compact}, \ie their initial algebras and
terminal coalgebras are isomorphic. For example, the interpretation
of~$\Sem{\tt{nat}}$ automatically contains the infinite
value~$\tt{Succ}(\tt{Succ}(\tt{Succ}(\dots)))$: it is the limit of~$\bot \leq
\tt{Succ}\,\bot \leq \tt{Succ}(\tt{Succ}\,\bot) \leq \cdots$.
We thus add a notion of \emph{totality}\footnote{Our notion seems unrelated to
intrinsic notions of totality that exist in effective domain theory.
\cite{berger:totality}} on top of the domains.
\begin{defi}\label{def:totality}\hfill \begin{enumerate}
    \item
      A \emph{domain with totality}~$(D,\T{D})$ is a domain~$D$
      together with a subset~$\T{D} \sub D$.

    \item
      An element of~$D$ is called \emph{total} when it belongs to~$\T{D}$.

    \item
      A function~$f$ from~$(D,\T{D})$ to~$(E,\T{E})$ is a function from~$D$
      to~$E$. It is \emph{total} if~$f(\T{D})\subseteq\T{E}$, \ie if it sends
      total elements to total elements.

    \item
      The category~$\Tot$ has domains with totality as objects and total
      continuous functions as morphisms.
  \end{enumerate}
\end{defi}
To interpret (co)datatypes inside the category~$\Tot$, it is enough to
describe the associated totality predicate. The following definition
corresponds to the natural interpretation of inductive / coinductive types
in the category of sets.
\begin{defi}\label{def:totality_in_types}
  Totality is defined by mutual induction on type definitions and type expressions.
  \begin{itemize}
    \item 
      If~$\tt T\in\mathcal{D}_r$ is a type definition of arity~$n$, and if~$\seq X$ is a
      sequence of~$n$ sets,
      \begin{itemize}
        \item
          totality for a datatype $\tt{T}_{\mu}$ is defined with
          ($\mu$ is the least fixed point operator)
          \[
          \T{T}(\seq X)
          \quad=\quad
          \mu X \ .
            \bigcup_{i=1,\dots,k} \Big\{ \C_i u\ \Bigm|\ u\in \T{S_i}(X, \seq X) \Big\}
          \]
          where~$\C_i
          : S_i \to T$ for~$i=1, \dots, k$ are the constructors for $\tt{T}_{\mu}$,

        \item
          totality for a codata type $\tt{T}_{\nu}$ is defined with
          ($\nu$ is the greatest fixed point operator)
          \[
          \T{T}(\seq X)
          \quad=\quad
          \nu X .
            \Big\{ \st{\D_1=u_1; \dots; \D_k=u_k} \ \Bigm|\ i=1,\dots,k \ \text{and}\  u_i\in \T{S_i}(X, \seq X) \Big\}
          \]
          where~$\tt{D}_i : T \to S_i$, $i=1,\dots,k$ are the destructors
          for~$\tt{T}_{\nu}$.
      \end{itemize}

    \item
      If~$T$ is a type expression,
      \begin{itemize}

        \item
          If $T=\x_i$ then~$\T{T}(\seq X) = X_i$

        \item
          If~$T=\tt T(\seq S)$, then~$\T{T}(\seq X) = \T{\tt T}\big(\T{S_1}(\seq X), \dots, \T{S_k}(\seq X)\big)$.

      \end{itemize}

  \end{itemize}
\end{defi}

Because these operators act on subsets of the set of all values and are
monotonic, the least and greatest fixed points exist by the Knaster-Tarski
theorem.
%\medbreak
% It is not difficult to see that each element of~$\T{T}$ is in~$\Sem{T}$ and
% since no element of~$\T{T}$ contains~$\bot$, $\T{T}$ contains only maximal
% elements of~$\Sem{T}$:
\begin{lem}\label{lem:total_max}
  If $T\in\mathcal{T}_r()$ is a closed type expression, then $\big(\Sem{T},
  \T{T}\big)$ is a domain with totality. Moreover, each~$t\in\T{T}$ is maximal
  in~$\Sem{T}$.
\end{lem}
\begin{proof}
Let~$v\in\T{T}(\seq X)$, we show that we can (coinductively) type it using the
rules from Definition~\ref{def:semantics_type}.

\begin{itemize}
  \item
    If~$T=\x_i$, then~$v\in\T{T}(\seq X) = X_i$, so that we can use rule~(2) to
    show that~$v\in\Sem{T}(\seq X)$.

  \item
    If~$T=\tt{T}_\mu(S_1, \dots)$, then~$\T{T}(\seq X) = \mu X . \bigcup_i
    \big\{ \C_i u\ |\ u\in \T{S_i}(X, \seq X) \big\}$, so that~$v$ is of the
    form~$\C_i u$ with~$u\in\T{S_i}(\tt{T}_\mu(\seq T), \seq X)$. We can use
    rule~(3) and continue coinductively.

  \item
    If~$T=\tt{T}_\nu(S_1, \dots)$, then~$\T{T}(\seq X) = \nu X . \big\{
    \st{\D_1=u_1; \dots; \D_k=u_k} \ \bigm|\ \forall i,u_i\in
    \T{S_i}(X, \seq X) \big\}$ so that~$v$ is of the form~$\st{\D_1=u_1; \dots;
    \D_k=u_k}$ where each~$u_i\in\T{S_i}\big(\T{T}(\seq X), \seq X\big)$.
    We can use rule~(4) and continue coinductively.

\end{itemize}
Note that we don't use the fact that~$\T{\tt{T}_\mu}$ is a least fixed point or
that~$\T{\tt{T}_\nu}$ is a greatest fixed point.

To show that elements of~$\T{T}$ are maximal, we need to show ``if~$u\leq v$
and~$u$ doesn't contain~$\bot$, then~$u=v$''. This is a trivial proof by
induction on~$u\leq v$.
\end{proof}
%%%>>>1

\subsection{Recursive Definitions} \label{sub:rec_def}%%%<<<1

Like in Haskell, recursive definitions are given by lists of
clauses. Here are two examples: the Ackermann function (using some syntactic
sugar for the constructors~\tt{Zero} and~\tt{Succ})
{\small\begin{alltt}
    val ack 0 n = n+1
      | ack (m+1) 0 = ack m 1
      | ack (m+1) (n+1) = ack m (ack (m+1) n)
\end{alltt}\label{def:ack}}\noindent
and the \tt{map} function on streams:
{\small\begin{alltt}
    val  map : ('a -> 'b) -> stream('a) -> stream('b)
       | map f \st{ Head = x ; Tail = s } = \st{ Head = f x ; Tail = map f s }
\end{alltt}}\noindent
Higher order parameters in recursive functions are allowed in the
implementation but they are ignored by the totality checker. Examples such as
\tt{map} above do use such higher order parameters but for simplicity's sake,
they are not formalized in the paper.\footnote{Note that we can't simply
ignore higher order parameters as they can hide some recursive calls:
{\small\begin{alltt}
  \hskip1cm   val app f x = f x       \hskip1cm--{\it non recursive}\\
  \hbox{}\hskip1cm   val g x = app g x \hskip1cm--{\it non terminating}
\end{alltt}}\noindent
The implementation first checks that all recursive functions are fully
applied. If that is not the case, the checker aborts and gives a negative
answer.}

\begin{defi}\label{def:definitions}
  A recursive definition is introduced by the keyword~\tt{val} and
  consists of a finite list of clauses of the form
  {\small\begin{alltt}
    | f \(p\sb1\) ... \(p\sb{n}\) = \(u\) \end{alltt}}\noindent
  where
  \begin{itemize}
    \item
      $\f$ is one of the function names being mutually defined,

    \item
      each~$p_i$ is a \emph{finite} \emph{linear} pattern
      \[
        p \qquad ::= \qquad
          \x_i \quad|\quad
          \C\ p \quad|\quad
          \st{ \D_1 = p_1 ; \dots ; \D_k = p_k }
      \]
      where each~$\x_i$ is a variable name that appear at most once in~$p$ (linearity),

    \item
      and~$u$ is a \emph{finite} term
      \[
        u \qquad ::= \qquad
          \x_i     \quad|\quad
          \C\ u \quad|\quad
          \st{ \D_1 = u_1 ; \dots ; \D_k = u_k } \quad | \quad
          \g\ u_1\ \dots\ u_k
      \]
      where~$k$ can be equal to~$0$, each~$\x_i$ is a variable name, and
      each~$\g$ is function name (recursive or otherwise).
  \end{itemize}
  The order of clauses is important and all clauses for a definition are
  grouped together.
\end{defi}

If a type is explicitly given, as in the definition of~\tt{map} above, the
definition is typed checked. Otherwise, as in the~\tt{ack} definition above,
the function's type is inferred. We'll always assume the definitions are
correct with respect to standard Hindley-Milner type
checking\label{rk:hindley_milner}~\cite{milner:polymorphism}. This includes in
particular checking that clauses of the definition cover all values of the
appropriate type and that no record is missing any field. Those steps are not
described in this paper but can be found in standard reference texts about the
implementation of functional programming languages~\cite[Section~8
and~9]{peyton:implementation}.

\subsubsection*{Standard semantics of a recursive definition} %%%<<<2
\label{sub:standard_semantics}

Hindley-Milner type checking guarantees that each list of clauses for
functions~$\f_1:T_1,\dots,\f_n:T_n$ (each~$T_i$ is a function type) gives rise
to an operator
\[
  \StdSem_{\f_1,\dots,\f_n} : \Sem{T_1}\times\dots\times\Sem{T_n} \to
  \Sem{T_1}\times\dots\times\Sem{T_n}
\]
where the semantics of types is extended with~$\Sem{T\to T'} = \big[\Sem{T}
\to \Sem{T'}\big]$.
The semantics of~$\f_1,\dots,\f_n$ is then defined as the fixed point of the
operator~$\StdSem_{\f_1,\dots,\f_n}$ which exists by Kleene
theorem.\footnote{The fixed point exists, but since domains contain the~$\bot$
value, the result can be a partial function.}
Let's describe more precisely the standard semantics of the definition in the
simple case of a single recursive function~\f taking a single argument. Given
an environment~$\rho$ for functions other than~\f, the recursive definition
for~$\f:A\to B$ gives rise to an operator~$\StdSem_{\rho,\f}$ on~$[\Sem{A}\to
\Sem{B}]$ called the ``standard semantics''. Its fixed point is the semantics
of~\f, written~$\Sem\f_\rho : \Sem{A} \to \Sem{B}$. The
operator~$\StdSem_{\rho,\f}$ is defined as follows.
\begin{defi}\label{def:matches}\leavevmode
  \begin{enumerate}
    \item Given a linear pattern~$p$ and a value~$v$, the unifier~$[p:=v]$ is the
      substitution defined inductively with
      \begin{itemize}
        \item
          $[\y := v] = [\y := v]$ where the RHS is the usual substitution
          of~\y by~$v$,

        \item
          $[\C p:= \C v] = [p:=v]$,

        \item
          $[\st{\D_1=p_1; \dots; \D_n=p_n} := \st{\D_1=v_1; \dots; \D_n=v_n}]
          \ =\ [p_1:=v_1] \cup \cdots \cup [p_n:=v_n]$, where the order of
          fields is irrelevant (note that because patterns are linear, the
          unifiers don't overlap),

        \item
          in all other cases, the unifier is undefined. Those cases are:

          \begin{itemize}
            \item
              $[\C p := \C' v]$ with~$\C\neq\C'$,

            \item
              $[\st{\dots}:=\st{\dots}]$ when the 2 records have different
              sets of fields,

            \item
              $[\C p := \st{\dots}]$ and $[\st{\dots}:= \C v]$.
          \end{itemize}
      \end{itemize}
      When the unifier~$[p:=v]$ is defined, we say that \emph{the value~$v$
      matches the pattern~$p$}.

    \item
      Given~$f:\Sem{A}\to\Sem{B}$ and $v\in\Sem{A}$,
      $\StdSem_{\rho,\f}(f)\big(v\big)$ can now be defined by:
      \begin{itemize}

        \item
          taking the first clause ``\tt{\f $p$ = $u$}'' in the definition of~\f
          where~$p$ matches~$v$,

        \item
          returning $\Sem{u[p:=v]}_{\rho,\f:=f}$.
      \end{itemize}
  \end{enumerate}
\end{defi}
An important property of Hindley-Milner type checking is that it ensures a
definition has a well defined semantics. In particular, there always is a
matching clause and the value ``$\bot$'' corresponds to non-termination, not
to failure of the evaluation mechanism (like projecting on a non-existing
field). However, it doesn't mean the definition is correct from a denotational
point of view. For that, we need it to be \emph{total} with respect to its
type. For example, the definition

{\small\begin{alltt}
    val all_nats : nat -> list(nat)
      | all_nats n = Cons n (all_nats (n+1))
\end{alltt}}
\noindent
is well typed and sends elements of the domain $\Sem{\tt{nat}}$ to the
domain~$\Sem{\tt{list}(\tt{nat})}$ but the image of \tt{Zero} is the infinite
list containing all the natural numbers. This is not total because totality
for~$\Sem{\tt{list}(\tt{nat})}$ contains only the finite lists.
Similarly, the definition
{\small\begin{alltt}
    val last_stream : stream(nat) -> nat
      | last_stream \st{Head=_; Tail=s} = last_stream s
\end{alltt}}
\noindent
sends any stream to~$\bot$, which is non total.
% Our aim is to describe a provably correct test that will detect such
% problems.
%%%>>>2

\subsubsection*{A note on projections} %%%<<<2

The syntax of definitions given in Definition~\ref{def:definitions} doesn't
allow projecting a record on one of its field. This makes the theory somewhat
simpler and doesn't change expressivity of the language because it is always
possible to rewrite a projection using one of the following techniques:
\begin{itemize}
  \item
    remove a projection on a previously defined function by introducing another
    function, as in
    {\small\begin{alltt}
      | f x = ... (g u).Fst ... \end{alltt}}\noindent
    being replaced by
    {\small\begin{alltt}
      | f x = ... projectFst (g u) ... \end{alltt}}\noindent
    where \tt{projectFst} is defined with
    {\small\begin{alltt}
      val projectFst \st{ Fst = x; Snd = y } = x \end{alltt}}

  \item
    remove a projection on a variable by extending the pattern on the left, as
    in
    {\small\begin{alltt}
      | f x = ... x.Head ... \end{alltt}}\noindent
    being replaced by
    {\small\begin{alltt}
      | f \st{ Head = h; Tail = t } = ... h ... \end{alltt}}

  \item
    remove a projection on the result of a recursively defined function by
    splitting the function into several mutually recursive functions, as in
    {\small\begin{alltt}
      | f : prod(\(\mathit{A}\), \(\mathit{B}\)) -> prod(\(\mathit{A}\), \(\mathit{B}\))
      | f p = ... (f u).Fst ... \end{alltt}}\noindent
    being replaced by
    {\small\begin{alltt}
      | f1 : prod(\(\mathit{A}\), \(\mathit{B}\)) -> \(\mathit{A}\)
      | f1 x = ... (f1 u1) ...
      | f2 : prod(\(\mathit{A}\), \(\mathit{B}\)) -> \(\mathit{B}\)
      ... \end{alltt}}\noindent
\end{itemize}
The first point is the simplest and most general but shouldn't be used to
remove projections on variables or recursive functions. Since the checker will
see each external function as a black box about which nothing is known,
introducing external functions in a recursive definition can hide information
and make totality checking harder.
Of course, the implementation of \chariot doesn't enforce this restriction and
the theory can be modified accordingly.
%%%>>>2

\subsubsection*{A subtle example}\label{sec:CounterExample} %%%<<<2

Here is an example showing that productivity and termination are not enough to
check validity of a recursive definition~\cite{altenkirch_danielsson:nested}.
We define the inductive type
{\small\begin{alltt}
    data stree where Node : stream(stree) -> stree
\end{alltt}}
\noindent
where the type of \tt{stream} was defined on page~\pageref{type:stream}. This
type is similar to the usual type of ``Rose trees'', but with streams instead
of lists. Because $\T{\tt{stream}}(\emptyset) = \emptyset$, we have that~$\mu
X.\T{\tt{streams}}(X) = \emptyset$. By Definition~\ref{def:totality_in_types},
~$\T{\tt{stree}}=\emptyset$: the interpretation of~\tt{stree} doesn't contain
any total value.
Consider however the following definitions:
{\small\begin{alltt}
    val bad_s : stream(stree)
      | bad_s = \st{ Head = Node bad_s ; Tail = bad_s }
\end{alltt}}
\noindent
This is well typed and guarded. Lazy evaluation of~\tt{bad\_s} or any of its
subterms terminates. The semantics of~\tt{bad\_s} doesn't contain~$\bot$ and
unfolding the definition gives
\begin{center}
  \begin{tikzpicture}[baseline=(current bounding box.center), %%%<<<3
       level distance=25pt,
       level 1/.style={sibling distance=180pt},
       level 2/.style={sibling distance=100pt},
       level 3/.style={sibling distance=60pt},
       level 4/.style={sibling distance=30pt},
       emptynode/.style={circle, inner sep=1.5pt, fill=black},
       treenode/.style={inner sep=5pt},
       font=\footnotesize,
       grow=down,
     ]
     \coordinate
       node[treenode] {$\tt{\st{Head=\BLANK ; Tail=\BLANK}}$}
         child {node[treenode] {$\tt{Node}$}
           child {node[treenode] {$\tt{\st{Head=\BLANK ; Tail=\BLANK}}$}
             child {node[treenode] {$\tt{Node}$}
               child {node[treenode] {\dots}}
             }
             child {node[treenode] {$\tt{\st{Head=\BLANK ; Tail=\BLANK}}$}
               child {node[treenode] {\dots}}
               child {node[treenode] {\dots}}
             }
           }
         }
         child {node[treenode] {$\tt{\st{Head=\BLANK ; Tail=\BLANK}}$}
           child {node[treenode] {$\tt{Node}$}
             child {node[treenode] {$\tt{\st{Head=\BLANK ; Tail=\BLANK}}$}
               child {node[treenode] {\dots}}
               child {node[treenode] {\dots}}
             }
           }
           child {node[treenode] {$\tt{\st{Head=\BLANK ; Tail=\BLANK}}$}
             child {node[treenode] {$\tt{Node}$}
               child {node[treenode] {\dots}}
             }
             child {node[treenode] {$\tt{\st{Head=\BLANK ; Tail=\BLANK}}$}
               child {node[treenode] {\dots}}
               child {node[treenode] {\dots}}
             }
           }
         }
       ;
  \end{tikzpicture}     %%%>>>3
\end{center}
Such a term clearly leads to inconsistencies. For example, the following
structurally decreasing function doesn't terminate when applied to~\tt{Node bad\_s}:
{\small\begin{alltt}
    val lower_left : stree -> empty
      | lower_left (Node \st{ Head = t; Tail = s }) = lower_left t
\end{alltt}}
\noindent
It is important to understand that \tt{lower\_left} is a total function and
that non termination of~\tt{lower\_left (Node bad\_s)} is a result
of~\tt{bad\_s} being non total.
%%%>>>1

%% vim600:set foldmarker=<<<,>>> foldmethod=marker fileencoding=ascii spelllang=en spell: %%

\section{Combinatorial Description of Totality}
\label{section:totality}

The set of total values for a given type can be rather complex when datatypes
and codatatypes are interleaved. Consider the following definition, given in
the introduction:
{\small\begin{alltt}\label{ex:inf}
    val inf = Node \st{ Left = inf; Right = inf }
\end{alltt}}
\noindent
This value can be total or not depending on its type:
\begin{itemize}
  \item 
    \tt{inf} of type \tt{tree} is {not total} with respect to the type
    definitions
    {\small\begin{alltt}
      codata pair('x,'y) where  Left : pair('x,'y) -> 'x
                              | Right : pair('x,'y) -> 'y
      data tree where Node : pair(tree, tree) -> tree     --{\it well-founded binary trees}
                    | Leaf : unit -> tree
    \end{alltt}}

  \item 
    \tt{inf} of type \tt{option(tree)} is {total} with respect to the type
    definitions
    {\small\begin{alltt}
      data option('x) where Node : 'x -> option('x)
                       | Leaf : unit -> option('x)
      codata tree where Left : tree -> option(tree)   --{\it non-well founded binary trees}
                       | Right : tree -> option(tree)
    \end{alltt}}
\end{itemize}
\noindent
It means in particular that the totality checker cannot be entirely untyped.
%%%>>>1

\subsection{Parity Games} %%%<<<1

A parity game is a two players game played on a finite directed simple graph
where each vertex, called \emph{node} in the sequel, is labeled by a natural
number called its \emph{priority}. We usually put labels on directed arcs
(called \emph{transitions} in the sequel), but besides making it easier to
refer to arcs, they don't really play any role.
% The \emph{height} of such a game is the maximum priority of its nodes.
%
When the node has odd priority, \emph{Marie} (or ``$\mu$'', or ``player'') is
required to play. When the node is even,\footnote{Assigning odd to one player
and even to the other is just a convention.} \emph{Nicole} (or ``$\nu$'', or
``opponent'') is required to play. A move is simply a choice of an outgoing
arc from the current node and the game continues from the new node.
When Nicole (or Marie) reaches a node without outgoing arcs, she looses. In
case of infinite play, the winning condition is
\begin{enumerate}
  \item
    if the maximal priority seen infinitely often is even, Marie wins,

  \item
    if the maximal priority seen infinitely often is odd, Nicole wins.
\end{enumerate}
\noindent
% Equivalently, the condition could be stated using the priorities of the
% transitions taken during the infinite play.
We will call a priority \emph{principal} if ``it is maximal among the
priorities appearing infinitely often''. The winning condition can thus be
rephrased as ``Marie wins an infinite play if and only if the principal
priority of the play is even''.

\smallbreak
In order to analyze types with parameters, we add parameter nodes~$\x_1$,
$\x_2$, \dots to the games. Those nodes have no outgoing transition and their
parity is, by convention,~$\infty$. On reaching them, Marie will be required
to choose
%\footnote{because~$\infty$ is obviously odd \smiley}
an element of some corresponding set~$X$ to finish the game. She'll win if she
can do it and loose if the set is empty.
Here are three examples of parity games:
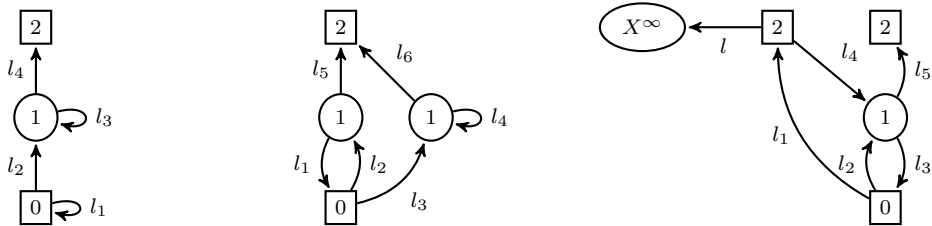
\[%%%<<<2
\label{ex:parity_games}
\begin{tikzpicture}[->,>=stealth',shorten >=1pt,auto,node distance=1.2cm,thick,
  mu node/.style={ellipse,draw,font=\nodeSize},
  nu node/.style={rectangle,draw,font=\nodeSize},
  every node/.style={font=\nodeSize},
  baseline=(one.base)
  ]
  \node[nu node] (one) {2};
  \node[mu node] (nat) [below of=one] {1};
  \node[nu node] (stream_nat) [below of=nat] {0};
  \path[every node/.style={font=\transitionSize}]
     (stream_nat) edge[loop right] node{$l_1$} (stream_nat)
     (stream_nat) edge node{$l_2$} (nat)
     (nat) edge[loop right] node{$l_3$} (nat)
     (nat) edge  node{$l_4$} (one)
     ;
\end{tikzpicture}
\mskip100mu
\begin{tikzpicture}[->,>=stealth',shorten >=1pt,auto,node distance=1.2cm,thick,
  mu node/.style={ellipse,draw,font=\nodeSize},
  nu node/.style={rectangle,draw,font=\nodeSize},
  every node/.style={font=\nodeSize},
  baseline=(one.base)
  ]
  \node[nu node] (one) {2};
  \node[mu node] (list_nat) [below of=one] {1};
  \node[mu node] (nat) [right of=list_nat] {1};
  \node[nu node] (nat_times_list_nat) [below of=list_nat] {0};

  \path[every node/.style={font=\transitionSize}]
  (nat_times_list_nat) edge[bend right] node[below right]{$l_3$} (nat)
  (nat_times_list_nat) edge[bend right] node[right]{$l_2$} (list_nat)
     (list_nat) edge node{$l_5$} (one)
     (list_nat) edge[bend right] node[left]{$l_1$} (nat_times_list_nat)
     (nat) edge node[above right]{$l_6$} (one)
     (nat) edge[loop right] node{$l_4$} (nat)
     ;
\end{tikzpicture}
\mskip50mu
% \begin{tikzpicture}[->,>=stealth',shorten >=1pt,auto,node distance=1.2cm,thick,
%   mu node/.style={ellipse,draw,font=\nodeSize},
%   nu node/.style={rectangle,draw,font=\nodeSize},
  % par node/.style={ellipse,draw,dotted,aspect=2,font=\nodeSize},
%   every node/.style={font=\nodeSize},
%   baseline=(one.base)
%   ]
%   \node[nu node] (one) {2};
%   \node[par node] [left of=one](X) {$X^\infty$};
%   \node[mu node] (binX) [below of=one] {1};
%   \node[nu node] (X_bin_bin) [below of=binX] {0};
%   \path[every node/.style={font=\transitionSize}]
%   (X_bin_bin) edge[bend left=45] node{$l_0$} (X)
%   (X_bin_bin) edge[min distance=1cm,bend right=80] node[right]{$l_3$} (binX)
%     (X_bin_bin) edge[bend right] node[right]{$l_2$} (binX)
%     (binX) edge[bend right] node[left]{$l_1$} (X_bin_bin)
%     (binX) edge node{$l_4$} (one)
%      ;
% \end{tikzpicture}
\begin{tikzpicture}[->,>=stealth',shorten >=1pt,auto,node distance=1.2cm,thick,
  mu node/.style={ellipse,draw,aspect=2,font=\nodeSize},
  nu node/.style={rectangle,draw,font=\nodeSize},
  par node/.style={ellipse,draw,dotted,aspect=2,font=\nodeSize},
  every node/.style={font=\nodeSize},
  baseline=(one.base)
  ]
  \node[nu node] (rtree) {2};
  \node[nu node] (one) [right=1cm of rtree]{2};
  \node[par node] (X) [left=1cm of rtree] {$\x_1^\infty$};
  \node[mu node] (list_rtree) [below of=one] {1};
  \node[nu node] (rtree_times_list_rtree) [below of=list_rtree] {0};

  \path[every node/.style={font=\transitionSize}]
     (rtree) edge node{$l$} (X)
     (rtree) edge node{$l_4$} (list_rtree)
     (list_rtree) edge[bend right] node[right]{$l_5$} (one)
     (list_rtree) edge[bend left] node[right]{$l_3$} (rtree_times_list_rtree)
     (rtree_times_list_rtree) edge[bend left] node[left]{$l_2$} (list_rtree)
     (rtree_times_list_rtree) edge[bend left] node[left]{$l_1$} (rtree)
     ;
\end{tikzpicture}
\]%%%>>>2

\bigbreak
\begin{defi}\label{def:semantic_games}
  Each position $p$ in a parity game~$G$ with parameters~$\x_1$, \dots, $\x_n$
  defines a functor~$||G||_p$ from~$\Set^n$
  to~$\Set$~\cite{santocanale:mu_bicomplete}. $||G||_p(\seq X)$  is defined by
  induction on the maximal finite priority of~$G$ and the number of positions
  with this priority:
  \begin{itemize}

    \item
      if all the positions are parameters, each position is interpreted by the
      corresponding parameter~$||G||_{\x_i}(\seq X) = X_i$;

    \item
      otherwise, take~$p$ to be one of the positions of maximal priority and
      construct~$G/p$ with a new parameter~$\x_0$ as follows: it is
      identical to~$G$, except that position~$p$ is replaced by~$\x_0$
      and all its outgoing transitions are removed.\footnote{This game is called
      the predecessor of~$G$~\cite{santocanale:mu_bicomplete}.}
      We define
      \begin{itemize}
        \item
          $||G||_p(\seq X) = \mu X. \big(||G/p||_{q_1}(X,\seq X) + \dots + ||G/p||_{q_k}(X, \seq X)\big)$
          if~$p$ had an odd priority and~$p \to q_1$, \dots $p\to q_k$ are all
          the transitions out of~$p$;

        \item
          $||G||_p(\seq X) = \nu X. \big(||G/p||_{q_1}(X, \seq X) \times \dots \times ||G/p||_{q_k}(X, \seq X)\big)$
          if~$p$ had an even priority and~$p \to q_1$, \dots $p\to q_k$ are
          all the transitions out of~$p$;

        \item
          $||G||_q(\seq X) = ||G/p||_q\big(||G||_p(\seq X),\seq X\big)$
          if~$p\neq q$.

      \end{itemize}
  \end{itemize}
\end{defi}
We give explicit names to nodes and write priorities as exponents.
Here is a small example to illustrate the construction of~$||G||$. Consider the
following parity game:
\[
G\quad=\quad
\begin{tikzpicture}[->,>=stealth',shorten >=1pt,auto,node distance=1.2cm,thick,
  mu node/.style={ellipse,draw,font=\nodeSize},
  nu node/.style={rectangle,draw,font=\nodeSize},
  every node/.style={font=\nodeSize},
  baseline=(current bounding box.center)
  ]
  \node[nu node] (one) {$a^2$};
  \node[mu node] (nat) [below of=one] {$b^1$};
  \path[every node/.style={font=\transitionSize}]
     (nat) edge[loop right] node{$s$} (nat)
     (nat) edge node{$t$} (one)
     ;
\end{tikzpicture}
\]
To compute~$||G||_a$ and~$||G||_b$, we need to compute~$||G/a||$, and
thus~$||G/a/b||$, given by:
\[
G/a \quad = \begin{tikzpicture}[->,>=stealth',shorten >=1pt,auto,node distance=1.2cm,thick,
  mu node/.style={ellipse,draw,font=\nodeSize},
  nu node/.style={rectangle,draw,font=\nodeSize},
  par node/.style={ellipse,draw,dotted,aspect=2,font=\nodeSize},
  every node/.style={font=\nodeSize},
  baseline=(current bounding box.center)
  ]
  \node[par node] (one) {$\x^\infty$};
  \node[mu node] (nat) [below of=one] {$b^1$};
  \path[every node/.style={font=\transitionSize}]
     (nat) edge[loop right] node{$s$} (nat)
     (nat) edge node{$t$} (one)
     ;
\end{tikzpicture}
\mskip100mu
  G/a/b \quad=
\begin{tikzpicture}[->,>=stealth',shorten >=1pt,auto,node distance=1.2cm,thick,
  mu node/.style={ellipse,draw,font=\nodeSize},
  nu node/.style={rectangle,draw,font=\nodeSize},
  par node/.style={ellipse,draw,dotted,aspect=2,font=\nodeSize},
  every node/.style={font=\nodeSize},
  baseline=(current bounding box.center)
  ]
  \node[par node] (one) {$\x^\infty$};
  \node[par node] (nat) [below of=one] {$\y^\infty$};
  \path[every node/.style={font=\transitionSize}]
     ;
\end{tikzpicture}
\]
By definition, $||G/a/b||_\y(X,Y) = Y$ and~$||G/a/b||_\x(X,Y) = X$. We thus
get the following
\begin{itemize}
  \item
    $B(X) := ||G/a||_b(X) = \mu_Y.\big(||G/a/b||_\x(X,Y) + ||G/a/b||_\y(X,Y)\big) = \mu_Y. (X+Y)$,

  \item
    $||G/a||_\x(X) = ||G/a/b||_\x(X,B(X)) = X$.
\end{itemize}
From that, we obtain
\begin{itemize}
  \item
    $A := ||G||_a = \nu_X.(\text{\emph{\scriptsize``empty product''}})$ as
    there is no outgoing transition from~$a$. This set is isomorphic
    to~$\mathbf{1}$, the one element set.

  \item
    $||G||_b = ||G/a||_b(A) = B(A) = \mu_Y.(Y+\mathbf{1}) \iso \mathbf{N}$.
    This set is isomorphic to the natural numbers.
\end{itemize}

\smallbreak
There is a strong link between the set~$||G||_p$ and the set of \emph{winning
strategies} for Marie in game~$G$ with initial position~$p$, where a strategy
is a kind of decision tree giving the ``next move'' on odd priority nodes.

\begin{defi}[Winning strategies]\label{def:strategy}
  Suppose~$G$ is a parity game with parameters~$\x_1$,\dots,~$\x_n$  and~$p$ a position
  in~$G$.
  %The functor~$\Win(G)_p:\Set^n\to\Set$ is defined as follows:
  Let~$\seq X=X_1,\dots,X_n$ be sets.
  \begin{enumerate}

    \item
      Write~$\mathcal{P}_p$ for the set of \emph{finite paths} starting
      from~$p$ in the graph underlying~$G$, equipped with the prefix
      order~$\sqsubseteq$.

    \item
      A subset~$S\subseteq \mathcal{P}_p$ is a \emph{pre-strategy} if:
      \begin{enumerate}
        \item
          it is downward closed: $\sigma_1 \sqsubseteq \sigma_2 \in S \implies \sigma_1\in S$,

        \item
          it is deterministic on odd nodes: if the last node of~$\sigma\in S$
          has odd priority, then there is a unique node~$v$ such
          that~$\sigma\cdot v \in S$,

        \item
          it is complete on even nodes:  if the last node~$u$ of~$\sigma\in S$
          has even priority, then for all neighbours~$v$ of~$u$, the
          path~$\sigma\cdot v$ is in~$S$.
      \end{enumerate}

    \item
      A \emph{leaf} in a pre-strategy~$S$ is the final node in a maximal finite
      path of~$S$.
      A \emph{strategy} is composed of a pre-strategy~$S$, together with a
      mapping from each parameter leaf~$\x_i$ of~$S$ to the corresponding
      set~$X_i$. (In particular,~$X_i$ cannot be empty.)

    \item
      A \emph{branch} in a pre-strategy~$S$ is a maximal, downward closed,
      linearly ordered subset of~$S$.
      A strategy is \emph{winning} if for all its infinite branches the
      maximal priority that appears infinitely often is
      even.
  \end{enumerate}
  We write~$\Win(G)_p(\seq X)$ for the set of such winning strategies for~$G$,
  from~$p$, with parameters~$\seq X$.

\end{defi}
Here are some simple examples where even node have out-degree~1, which means
strategies are just paths in the underlying graph. Note that paths are defined
as sequences of consecutive nodes but we sometimes give them using sequences
of transitions for readability.
\[
\begin{tikzpicture}[->,>=stealth',shorten >=1pt,auto,node distance=1.2cm,thick,
  mu node/.style={ellipse,draw,font=\nodeSize},
  nu node/.style={rectangle,draw,font=\nodeSize},
  every node/.style={font=\nodeSize},
  baseline=(current bounding box.center)
  ]
  \node[nu node] (one) {$a^2$};
  \node[mu node] (nat) [below of=one] {$b^1$};
  \path[every node/.style={font=\transitionSize}]
     (nat) edge[loop right] node{$s$} (nat)
     (nat) edge node{$t$} (one)
     ;
\end{tikzpicture}
\mskip100mu
\begin{tikzpicture}[->,>=stealth',shorten >=1pt,auto,node distance=1.2cm,thick,
  mu node/.style={ellipse,draw,font=\nodeSize},
  nu node/.style={rectangle,draw,font=\nodeSize},
  every node/.style={font=\nodeSize},
  baseline=(current bounding box.center)
  ]
  \node[mu node] (c) {$c^1$};
  \node[nu node] (d) [above left=1cm of c] {$d^2$};
  \node[nu node] (e) [above right=1cm of c] {$e^2$};
  \path[every node/.style={font=\transitionSize}]
     (c) edge[bend right] (d)
     (d) edge[bend right] (c)
     (c) edge[bend right] (e)
     (e) edge[bend right] (c)
     ;
\end{tikzpicture}
\mskip100mu
\begin{tikzpicture}[->,>=stealth',shorten >=1pt,auto,node distance=1.2cm,thick,
  mu node/.style={ellipse,draw,font=\nodeSize},
  nu node/.style={rectangle,draw,font=\nodeSize},
  every node/.style={font=\nodeSize},
  baseline=(current bounding box.center)
  ]
  \node[nu node] (u) {$u^2$};
  \node[mu node] (t) [below left of=u] {$t^1$};
  \node[nu node] (s) [below left of=t] {$s^0$};
  \path[every node/.style={font=\transitionSize}]
     (u) edge[bend right] (t)
     (t) edge[bend right] (u)
     (t) edge[bend right] (s)
     (s) edge[bend right] (t)
     ;
\end{tikzpicture}
\]
In the first game, starting from~$b$, a strategy is either of the
form~$(s^{n}t)^\downarrow = \{\varepsilon, s, ss, \dots, s^n, s^n t\}$, or
equal to~$s^{\infty\downarrow} = \{\varepsilon, s, ss, \dots\}$. This infinite
strategy isn't winning because the maximal priority visited infinitely often
is odd: $1$.
For the middle game, there are many infinite strategies from~$c$ that can
alternate between the~$d$ and~$e$ node arbitrarily. All of them are winning 
% $(st)^{\infty\downarrow}=\{\varepsilon, s, st, sts, stst, ststs, \dots \}$,
as the maximal priority visited infinitely often is always even: $2$. For the
last game, we can define both winning and non-winning infinite strategies
starting from~$t$.

\medbreak
An important result is:
\begin{propC}[{\cite[Theorem~5.4]{santocanale:mu_bicomplete}}]\label{prop:muBicomplete}\leavevmode
      There is a natural isomorphism
      \[
      ||G||_p \iso \Win(G)_p
      \ .
      \]
\end{propC}
%%%>>>1

\subsection{Parity Games from Types} %%%<<<1

We can construct a parity game~$G$ from any type~$T$ in such a way that~$\T{T}
\iso ||G||_T$, for some distinguished position~$T$ in~$G$.

\begin{defi}\label{def:types_LTS}\leavevmode
  \begin{enumerate}
    \item
      Given a (fixed) list of type definitions, we consider the following
      transition system:
      \begin{itemize}
        \item
          nodes are type expressions, possibly with parameters,

        \item
          transitions are labeled by constructors and destructors: a
          transitions~$T_1 \transition{t} T_2$ is either a destructor~$t$ of
          type~$T_1 \to T_2$ or a constructor~$t$ of type~$T_2 \to T_1$ (note the
          reversal).
      \end{itemize}

    \item
      If~$T$ is a type expression, the \emph{graph of~$T$} is defined as the
      part of the above transition system that is reachable from~$T$.
  \end{enumerate}
\end{defi}
Here is for example the graph of \tt{list(nat)}
%(this game is the second example from page~\pageref{ex:parity_games}):
\[
\begin{tikzpicture}[->,>=stealth',shorten >=1pt,auto,node distance=1.5cm,thick,
  mu node/.style={ellipse,draw,aspect=2,font=\nodeSize},
  nu node/.style={rectangle,draw,font=\nodeSize},
  every node/.style={font=\nodeSize},
  baseline=(one.base),
  ]
  \node (one) {$\tt{unit}$};
  \node (list_nat) [below of=one] {$\tt{list(nat)}$};
  \node (nat) [right=1cm of list_nat] {$\tt{nat}$};
  \node (nat_times_list_nat) [below of=list_nat] {$\tt{prod(nat,list(nat))}$};

  \path[every node/.style={font=\transitionSize}]
  (nat_times_list_nat) edge[bend right] node[below right]{\tt{Fst}} (nat)
  (nat_times_list_nat) edge[bend right] node[right]{\tt{Snd}} (list_nat)
  (list_nat) edge node[left]{\tt{Nil}} (one)
     (list_nat) edge[bend right] node[left]{\tt{Cons}} (nat_times_list_nat)
     (nat) edge[bend right] node[above right]{\tt{Zero}} (one)
     (nat) edge[loop right] node[right]{\tt{Succ}} (nat)
     ;
\end{tikzpicture}
\]
The transition system is set up so that
\begin{itemize}
  \item
    on data nodes, a transition is a choice of constructor for the origin
    type,

  \item
    on codata nodes, a transition is a choice of field for a record for the
    origin type.
\end{itemize}
Because of that, Hindley-Milner type checking will ensure that a value of
type~$T$ gives a strategy for a game on the graph of~$T$ where Marie (the
player) chooses constructors and Nicole (the opponent) chooses destructors
(that will be Lemma~\ref{lem:terms_strategies}). We will, in
Definition~\ref{def:type_game}, add priorities so that
\begin{itemize}

  \item
    datatype nodes are odd and codatatype nodes are even,

  \item
    the order of priorities correspond in a precise way to the interleaving of
    least and greatest fixed points.
\end{itemize}
Checking that this strategy is \emph{winning} will be the goal of the totality
checker.

\medbreak
Note that when some of the types have parameters, the transition system is
infinite: it will for example contain~\tt{list('x)}, \tt{list(list('x))},
\tt{list(list(list('x)))}, etc. However, we have
\begin{lem}\label{lem:finite}
  For any type~$T$, the graph of~$T$ is finite.
\end{lem}
\noindent
This relies on the fact that recursive types are uniform: their parameters are
constant in their definition. It becomes false if we were to allow more
general types like
{\small\begin{alltt}
    data t('x) where
      | Empty : unit              -> t('x)
      | Cons : prod('x, t(\underline{t(\char13x)})) -> t('x)   -- !!! not uniform
\end{alltt}}
\noindent
The graph of \tt{t(nat)} would contain the following infinite chain:
\[
\begin{tikzpicture}[->,>=stealth',shorten >=1pt,auto,node distance=1.5cm,thick,
  mu node/.style={ellipse,draw,aspect=2,font=\nodeSize},
  nu node/.style={rectangle,draw,font=\nodeSize},
  every node/.style={font=\nodeSize},
  ]
  \node (t) {$\tt{t(nat)}$};
  \node (pt) [right=.5cm of t] {$\tt{prod(nat,t(t(nat)))}$};
  \node (tt) [right=.5cm of pt] {$\tt{t(t(nat))}$};
  \node (ptt) [right=.5cm of tt] {$\tt{prod(t(nat),t(t(t(nat))))}$};
  \node (ttt) [right=.5cm of ptt] {$\tt{...}$};
  \path[every node/.style={font=\transitionSize}]
  (t) edge[bend left] node[above]{\tt{Cons}} (pt)
  (pt) edge[bend left] node[above]{\tt{Snd}} (tt)
  (tt) edge[bend left] node[above]{\tt{Cons}} (ptt)
  (ptt) edge[bend left] node[above]{\tt{Snd}} (ttt);
\end{tikzpicture}
\]
Before proving the lemma, the following definition will be useful.
\begin{defi}
  Write $T_1\sqsubseteq T_2$ if~$T_1$ is a subexpression of~$T_2$.
  More precisely:
  \begin{itemize}
    \item
      $T \sqsubseteq X$ iff $T = X$,

    \item
      $T \sqsubseteq \tt{T}(T_1,\dots,T_n)$ if and only if $T =
      \tt{T}(T_1,\dots,T_n)$ or $T\sqsubseteq T_1$ or \dots\ or $T\sqsubseteq
      T_n$.
  \end{itemize}
\end{defi}
\begin{proof}[Proof of Lemma~\ref{lem:finite}]
  From Definition~\ref{def:type_defs}, we know that the definition of a
  (co)datatype only uses parameters and type expressions of strictly smaller
  rank. Moreover, in the presence of mutual type definitions, two types of the
  same order are part of the same mutual definition.

  Suppose by contradiction that~$T$ is a type expression of minimal head rank
  (\cf Definition~\ref{def:type_defs}) with an infinite graph.
  Since the graph of~$T$ has finite out-degree, K\"onig's lemma implies it
  contains an infinite path~$\rho = T \to T_1 \to T_2 \to \cdots$ without
  repeated vertex.

  There must be infinitely many~$T_i$ with head rank equal to~$\kappa$.
  Otherwise, we could construct an infinite path with a strictly smaller head
  rank: take the first~$T_{n}$ s.t. all later~$T_{m}$ have head
  rank~$\lambda<\kappa$ and let~$T'_m$, for~$m>n$ be obtained from~$T_m$ by
  replacing all subexpressions of rank~$\kappa$ by a new parameter~\x.

  Because no~$T_m$ has a head rank~$\kappa$, any transition~$T_m \to
  T_{m+1}$ is also a transition~$T'_m \to T'_{m+1}$ and we obtain an infinite
  path~$T'_n\to T'_{n+1}\to\cdots$ with rank strictly smaller than~$\kappa$.

  But if~$T$ has rank~$\kappa$, its subexpressions of head rank~$\kappa$ have
  fixed arguments by uniformity. So all subexpressions of head rank~$\kappa$
  along the infinite path must be subexpressions of~$T_0$. There are only
  finitely many of those, contradicting the previous remark.
\end{proof}

\begin{defi}\label{def:type_game}\leavevmode
  If~$T$ is a type expression, a \emph{parity game for~$T$} is a parity game
  on the graph of~$T$ (Definition~\ref{def:types_LTS}) which satisfies the
  following conditions:
  \begin{enumerate}
    \item
    if~$T_0$ is a datatype, its priority is odd,

    \item
    if~$T_0$ is a codatatype, its priority is even,

    \item
    if~$T_1\sqsubseteq T_2$, then the priority of~$T_1$ is greater than the priority of~$T_2$.
  \end{enumerate}
\end{defi}
\begin{lem}\label{lem:type_game}
  Each type expression has a parity game.
\end{lem}
\begin{proof}
The relation~$\sqsubseteq$ is a strict order and doesn't contain cycles. Its
restriction to the graph of~$T$ can be linearized. This gives the relative
priorities of the nodes and ensures condition~(5) from the definition.
Starting from the least priorities, we can now choose a priority odd / even
compatible with this linearization.
\end{proof}

We don't actually need to linearize the graph and can instead chose
a \emph{normalized parity game}, \ie one that minimizes gaps in priorities.
Here are the first two parity games from page~\pageref{ex:parity_games},
seen as parity games for \tt{stream(nat)} and \tt{list(nat)}. Priorities are
written as exponents and their parity can be seen in the shape (square or
round) of nodes.
% The third parity game corresponds to the type~$\tt{bin}(X)$ of binary trees
% labeled by~$X$ and is left as an exercise.
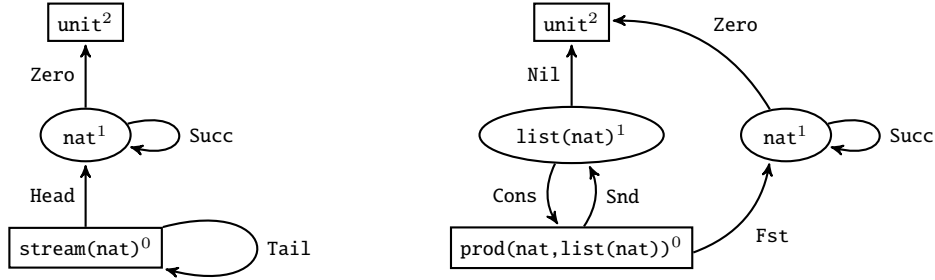
\[%%%<<<2
\begin{tikzpicture}[->,>=stealth',shorten >=1pt,auto,node distance=1.5cm,thick,
  mu node/.style={ellipse,draw,aspect=2,font=\nodeSize},
  nu node/.style={rectangle,draw,font=\nodeSize},
  every node/.style={font=\nodeSize},
  baseline=(one.base)
  ]
  \node[nu node] (one) {$\tt{unit}^2$};
  \node[mu node] (nat) [below of=one] {$\tt{nat}^1$};
  \node[nu node] (stream_nat) [below of=nat] {$\tt{stream(nat)}^0$};
  \path[every node/.style={font=\transitionSize}]
  (stream_nat) edge node {\tt{Head}} (nat)
  (stream_nat) edge [loop right] node{\tt{Tail}} (stream_nat)
  (nat) edge node{\tt{Zero}} (one)
  (nat) edge [loop right] node{\tt{Succ}} (nat)
  ;
\end{tikzpicture}
\mskip80mu
\begin{tikzpicture}[->,>=stealth',shorten >=1pt,auto,node distance=1.5cm,thick,
  mu node/.style={ellipse,draw,aspect=2,font=\nodeSize},
  nu node/.style={rectangle,draw,font=\nodeSize},
  every node/.style={font=\nodeSize},
  baseline=(one.base)
  ]
  \node[nu node] (one) {$\tt{unit}^4$};
  \node[mu node] (list_nat) [below of=one] {$\tt{list(nat)}^1$};
  \node[mu node] (nat) [right=1cm of list_nat] {$\tt{nat}^3$};
  \node[nu node] (nat_times_list_nat) [below of=list_nat] {$\tt{prod(nat,list(nat))}^0$};

  \path[every node/.style={font=\transitionSize}]
  (nat_times_list_nat) edge[bend right] node[below right]{\tt{Fst}} (nat)
  (nat_times_list_nat) edge[bend right] node[right]{\tt{Snd}} (list_nat)
  (list_nat) edge node[left]{\tt{Nil}} (one)
  (list_nat) edge[bend right] node[left]{\tt{Cons}} (nat_times_list_nat)
  (nat) edge[bend right] node[above right]{\tt{Zero}} (one)
  (nat) edge[loop right] node[right]{\tt{Succ}} (nat)
  ;
\end{tikzpicture}
% \mskip50mu
% \begin{tikzpicture}[->,>=stealth',shorten >=1pt,auto,node distance=1.5cm,thick,
%   mu node/.style={ellipse,draw,aspect=2,font=\nodeSize},
%   nu node/.style={rectangle,draw,font=\nodeSize},
% every node/.style={font=\nodeSize},
%  baseline=(one.base)
%   ]
%   \node[nu node] (one) {2};
%   \node [left of=one](X) {$X$};
%   \node[mu node] (binX) [below of=one] {1};
%   \node[nu node] (X_bin_bin) [below of=binX] {0};
%   \path[every node/.style={font=\transitionSize}]
%     (X_bin_bin) edge [bend left=35] (X)
%     (X_bin_bin) edge [bend right=65] (binX)
%     (X_bin_bin) edge [bend right](binX)
%     (binX) edge [bend right] (X_bin_bin)
%     (binX) edge (one)
%      ;
% \end{tikzpicture}
\]%%%>>>2
The last example from page~\pageref{ex:parity_games} corresponds to
a coinductive version of Rose trees:
{\small\begin{alltt}
  codata rtree('x) where
  | Root : rtree('x) -> 'x
  | Subtrees : rtree('x) -> list(rtree('x))
\end{alltt}}
\noindent
with parity game
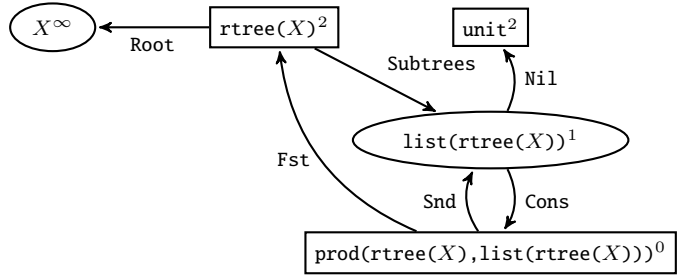
\[%%%<<<2
\begin{tikzpicture}[->,>=stealth',shorten >=1pt,auto,node distance=1.5cm,thick,
  mu node/.style={ellipse,draw,aspect=2,font=\nodeSize},
  nu node/.style={rectangle,draw,font=\nodeSize},
  par node/.style={ellipse,draw,dotted,aspect=2,font=\nodeSize},
  every node/.style={font=\nodeSize},
  baseline=(one.base)
  ]
  \node[nu node] (rtree) {$\tt{rtree(x)}^2$};
  \node[nu node] (one) [right=1.5cm of rtree]{$\tt{unit}^2$};
  \node[par node] (X) [left=1.5cm of rtree] {$\x^\infty$};
  \node[mu node] (list_rtree) [below of=one] {$\tt{list(rtree(x))}^1$};
  \node[nu node] (rtree_times_list_rtree) [below of=list_rtree]
  {$\tt{prod(rtree(x),list(rtree(x)))}^0$};

  \path[every node/.style={font=\transitionSize}]
  (rtree) edge node{\tt{Root}} (X)
  (rtree) edge node{\tt{Subtrees}} (list_rtree)
  (list_rtree) edge[bend right] node[right]{\tt{Nil}} (one)
  (list_rtree) edge[bend left] node[right]{\tt{Cons}} (rtree_times_list_rtree)
  (rtree_times_list_rtree) edge[bend left] node[left]{\tt{Snd}} (list_rtree)
  (rtree_times_list_rtree) edge[bend left] node[left]{\tt{Fst}} (rtree)
  ;
\end{tikzpicture}
\]%%%>>>2
% The node~$\tt{list}(\tt{rtree}(X))$ is necessarily buried somewhere
% between~$\tt{rtree}(X)$ and~$\tt{prod}(\tt{rtree}(X),\tt{list}(\tt{rtree}(X)))$.
%
As the examples show, the priority of a type can be minimal
(\tt{stream(nat)\p0}), maximal (\tt{rtree($X$)\p2}) or somewhere in between
(\tt{list(nat)\p1}) in its parity game.

\bigbreak
The semantics of a parity game (Definition~\ref{def:semantic_games}) and that
of the totality semantics of a type (Definition~\ref{def:totality_in_types})
are similar in that they interleave greatest and least fixed points. Parity
games of types are designed to get the following.
\begin{prop}\label{prop:par_game_iso}\leavevmode
  For any type parity game~$G$ and for any node~$T$ in~$G$, we have~$||G||_T
  \iso \T{T}$.
\end{prop}
\begin{proof}
We prove ``\emph{for any type expression~$T$ and parity game~$G$ containing a
parity game for~$T$ as a simply connected component, we have~$||G||_T \iso
\T{T}$}'' by induction on~$G$.

To simplify the argument, we assume that all non-recursive constructors [resp.
destructors] for type~$\tt T(\seq\x)$ are of type~$\x_i\to\tt{T(\seq\x)}$
[resp.~$\tt{T(\seq\x)}\to\x_i$]. Every type definition can be transformed into
one such by adding additional parameters. For example, we can
replace~$\tt{list(\x)}$ by
{\small\begin{alltt}
  data list2('x, 'y) where  Nil  : 'y                      -> list2('x, 'y)
  | Cons : prod('x, list2('x, 'y)) -> list2('x, 'y)
\end{alltt}}\noindent
and the old ``\tt{list(x)}'' is the same as the new ``\tt{list2(x, unit)}'': it
has the same semantics, totality and parity game.

% \[%%%<<<2
% \begin{tikzpicture}[->,>=stealth',shorten >=1pt,auto,node distance=1.5cm,thick,
%   mu node/.style={ellipse,draw,aspect=2,font=\nodeSize},
%   nu node/.style={rectangle,draw,font=\nodeSize},
%   par node/.style={ellipse,draw,dotted,aspect=2,font=\nodeSize},
%   every node/.style={font=\nodeSize},
%   baseline=(one.base)
%   ]
%   \node[par node] (rtree) {$\T{\tt{rtree($X$)}}^\infty$};
%   \node[nu node] (one) [right=1.5cm of rtree]{$\tt{unit}^2$};
%   \node[par node] (X) [left=1.5cm of rtree] {$X^\infty$};
%   \node[mu node] (list_rtree) [below of=one] {$\tt{list(rtree($X$))}^1$};
%   \node[nu node] (rtree_times_list_rtree) [below of=list_rtree]
%   {$\tt{prod(rtree($X$),list(rtree($X$)))}^0$};

%   \path[every node/.style={font=\transitionSize}]
%      (list_rtree) edge[bend right] node[right]{\tt{Nil}} (one)
%      (list_rtree) edge[bend left] node[right]{\tt{Cons}} (rtree_times_list_rtree)
%      (rtree_times_list_rtree) edge[bend left] node[left]{\tt{Snd}} (list_rtree)
%      (rtree_times_list_rtree) edge[bend left] node[left]{\tt{Fst}} (rtree)
%      ;
% \end{tikzpicture}
% \]%%%>>>2

\begin{itemize}

  \item
  The result is obvious when all the nodes of~$G$ have
  priority~$\infty$:~$T$ is necessarily a parameter~$\x_i$ and we have
  $||G||_{\x_i}(\seq X) = X_i = \T{\x_i}$.

  \item
  If~$T$ is a position of maximal finite priority in~$G$ and~$T=\tt T_\mu(\seq T)$
    starts with a datatype with constructors~$\C_i:S_i\to T$, $i=1, \dots,k$.
    We have
    \[\begin{array}{rcll}
    ||G||_T(\seq X)
      &=&
    \mu X.\big(
      ||G/T||_{S_1}(X,\seq X)
      +\cdots+
      ||G/T||_{S_k}(X,\seq X)
    \big)
      & \text{\footnotesize(Definition~\ref{def:semantic_games})}\\
      &\iso&
    \mu X.\big(
      |S_1|(X,\seq X)
      +\cdots+
      |S_k|(X,\seq X)
    \big)
      & \text{\footnotesize(induction hypothesis)}\\
      &\iso&
    \mu X.\big(
      \C_1|S_1|(X,\seq X)
      \cup\cdots\cup
      \C_k|S_k|(X,\seq X)
    \big)
    \\
      &=&
    |T|
      & \text{\footnotesize(Definition~\ref{def:totality_in_types})}
    \end{array}\]
      The reason we can apply the induction hypothesis on~$G/T$ is that it
      contains the parity games for each~$S_i$: because types have a special
      form, transition can only be of 2 forms:
      \begin{itemize}
        \item
          non recursive constructor / destructor: $\tt{S}(\seq S) \to S_i$
          where the priority increases,

        \item
          recursive constructor / destructor: $\tt{S}(\seq S) \to R$
          where~$\tt{S}(\seq S)$ is a subexpression of~$R$, \ie the priority
          decreases.
      \end{itemize}
      Because~$T$ is of maximal finite priority, the only possible
      transitions to~$T$ are of the first kind. In~$G/T$, which contains a
      new parameter~\x, they become
      transitions~$\tt{S}(\seq\x)\to\x_i$.\footnote{Without our
      hypothesis,~$\tt{Nil}:\tt{list(nat)}\to\tt{unit}$ could be replaced
      by~$\tt{Nil}:\tt{list(nat)}\to\x$ in~$G/{\tt{unit}}$ so that~$G/\tt{unit}$
      doesn't really contain a parity game for~\tt{list(nat)}.} This implies
      that~$G/T$ contains a type parity game for~$S_i$.

    \item
      The reasoning is similar if $T=\tt T_\nu(\seq X)$ is a codatatype.

    \item
      if the priority of~$T$ is not maximal, write~$M$ for the position of
      maximal finite priority. We have
      \[\begin{array}{rcll}
        ||G||_T(\seq X)
        &=&
        ||G/M||_T\big(||G||_M(\seq X), \seq X\big)
        & \text{\footnotesize(Definition~\ref{def:semantic_games})}\\
        &\iso&
        ||G/M||_T\big(\T{M}(\seq X), \seq X\big)
        & \text{\footnotesize(previous point)}\\
        &\iso&
        |T|(\seq X)
        & \text{\footnotesize(induction hypothesis, (*))}
      \end{array}\]

      The last isomorphism is the most subtle: $G/M$ contains a new
      parameter~\x
      corresponding to~$M$.
      Like above, any transition to~$M$ in~$G$ is transformed into a
      transition to the new parameter~$M$ in~$G/M$.
      By induction hypothesis, we get that~$||G/M||_T(\x,\seq X)
      \iso \T{T'}(\x, \seq X)$ where~$T'$ is the type expression~$T$ where
      every occurrence of~$M$ has been replaced by~\x. The last step decomposes into
      \[\begin{array}{rcll}
        ||G/M||_T\big(\T{M}(\seq X),\seq X\big)
        &\iso&
        \T{T'}\big(\T{M}(\seq X),\seq X\big)
        \\
        &\iso&
        |T|(\seq X)
        & \text{\footnotesize(definition of totality)}
      \end{array}\]
      Indeed, on reaching~\x when computing~$\T{T'}(\T{M},\seq X)$, we
      return~$M$; just like on reaching~$M$ when computing~$\T{T}(\seq
      X)$. \qedhere
\end{itemize}
\end{proof}
%%%>>>1

\subsection{Strategies from Terms} %%%<<<1

Strategies~(Definition~\ref{def:strategy}) are defined as order-theoretic
trees. Because of the way types parity games are defined, they are equivalent
to values in the corresponding type.
\begin{lem}\label{lem:terms_strategies}
  For any type~$T$, and associated parity game~$G$, the set of strategies
  for~$G$ starting from node~$T$ is isomorphic to the set of maximal elements
  in ~$\Sem{T}$.
\end{lem}
\begin{proof}
  Let~$t$ be a maximal element in~$\Sem{T}$, that is, an element of~$\Sem{T}$
  which doesn't contain~$\bot$. By definition, each finite branch of a maximal
  element of~$\Sem{T}$ is a finite path from~$T$ in the graph of~$T$.
  \begin{itemize}
    \item
      If~$T$ is a datatype/odd position, $t$ chooses precisely one
      constructor. The set of finite branches of~$t$ is thus
      \emph{deterministic} on odd positions.

    \item
      If~$T$ is a codatatype/even position, $t$ contains one field for each
      destructor. The set of finite branches of~$t$ is thus \emph{complete} on
      even positions.
  \end{itemize}

  \smallbreak
  Conversely, we can construct a maximal element~$\widehat{s}$ of~$\Sem{T}$ from
  any strategy~$s$ from~$T$ in~$G$ coinductively.
  \begin{itemize}
    \item
      If~$T$ is a data, by determinism, all non-empty paths of~$s$ start with
      the same constructor~$\C$: this is the head constructor of~$\widehat{s}$
      and we continue by considering the strategy~$s/\C$ obtained from~$s$ by
      removing the head constructor of each of its paths: $\widehat{s} := \C\
      \widehat{s/\C}$.

    \item
      If~$T$ is a codata, by completeness, there are paths starting with each
      destructor~$\D_k$ of~$T$. In that case, we can put~$\widehat{s} =
      \st{\dots; \D_k = \widehat{s/\D_k}; \dots}$.

  \end{itemize}
  This construction works because by construction, $s/\C$ are strategies for
  the appropriate types.
\end{proof}

\medbreak
Putting together Proposition~\ref{prop:par_game_iso},
Proposition~\ref{prop:muBicomplete}
and Lemma~\ref{lem:terms_strategies} we finally get
\begin{cor}\label{cor:totality_values}
  If $T$ is a type and~$G$ a parity game for~$T$, we have
  % a natural isomorphism
  $\Win(G)_T \iso \T{T}$.
  In particular, $v\in\Sem{T}$ is total iff every branch of~$v$ has even
  principal priority.
\end{cor}
The only thing to note is that priorities are not part
of the value~$v\in\Sem{T}$ but are read in~$G$.
As a final example in this section, let's consider the problematic example
from page~\pageref{sec:CounterExample}. The parity game for~\tt{stree} is given
by
\[
\begin{tikzpicture}[->,>=stealth',shorten >=1pt,auto,node distance=1.2cm,thick,
  mu node/.style={ellipse,draw,font=\nodeSize},
  nu node/.style={rectangle,draw,font=\nodeSize},
  every node/.style={font=\nodeSize},
  ]
  \node[mu node] (stree) {\tt{stree}\p1};
  \node[nu node] (stream_stree) [below=1cm of stree] {\tt{stream(stree)}\p0};
  \path[every node/.style={font=\transitionSize}]
     ($(stream_stree)+(1cm,-2mm)$) edge[loop, distance=2cm, out=-30,in=30,
     shorten <= .2cm,shorten >=.2cm] node[right]{\tt{Tail}}
     ($(stream_stree)+(1cm,2mm)$)
     % (stream_stree) edge[loop right] node{\tt{Tail}} (stream_stree)
     (stream_stree) edge[bend right] node[right]{\tt{Head}} (stree)
     (stree) edge[bend right] node[left]{\tt{Node}} (stream_stree)
     ;
\end{tikzpicture}
\]
There is a single strategy from the lower node of this parity game and it
obviously contains an infinite branch with principal priority~1. This strategy
cannot be winning, \ie the corresponding value cannot be total.

%%%>>>1

\subsection{Forgetting Types} %%%<<<1

If a term~$t$ in \chariot (not necessarily a value) is of type~$T$, it will
generate a strategy in~$G$, a parity game for~$T$. Thanks to
Corollary~\ref{cor:totality_values}, the definition of~$t$ is total if and only
if the corresponding strategy is winning.\footnote{Usually, $t$ will be a
function, resulting in some additional complexity.}

In order to talk about priorities in \chariot, we annotate each occurrence of
constructor / destructor in a definition with its priority taken from one of
the type's parity game. This can be done during Hindley-Milner type checking:
\begin{itemize}
  \item
    each instance of a constructor / destructor is annotated by its type
    during type checking,

  \item
    all the types appearing in the definitions are gathered (and completed) to
    a parity game,

  \item
    each constructor / destructor is then given the priority of its
    type.
\end{itemize}
Once type checking is done, the type of constructors / destructors can be
erased and only their priorities are kept.
We end up with definitions like
{\small\begin{alltt}
    val length : list\p1(x) -> nat\p1
      | length Nil\p1 = Zero\p1
      | length (Cons\p1\st{Fst\p0=_; Snd\p0=l}) = Succ\p1 (length l)
\end{alltt}}\noindent
Note that priorities are inferred and only used while checking totality.
They are never shown to the end user.

We thus refine the notion of value from the previous
section by adding priorities on constructors and destructors.
\begin{defi}\label{def:refined_values}
  The set of \emph{values with leaves in~$X_1$,\dots, $X_n$},
  written~$\V(X_1,\dots, X_n)$ is defined coinductively by the grammar
  \[
    v \qquad::=\qquad
    \bot \quad|\quad
    x \quad|\quad
    \C\p p\,v \quad|\quad
    \st{\D_1=v_1; \dots; \D_k=v_k}\p p
  \]
  where
  \begin{itemize}
    \item
      each $x$ is in one of the~$X_i$,

    \item
      each priority $p$ belong to a finite set of natural numbers,

    \item
      each $\C$ belongs to a finite set of \emph{constructors}, and their
      priority is odd,

    \item
      each $\tt{D}_i$ belongs to a finite set of \emph{destructors}, and their
      priority is even,

    \item
      $k$ can be 0.
  \end{itemize}
\end{defi}
Corollary~\ref{cor:totality_values} gives an intrinsic notion of totality
on~$\V$.
\begin{defi}\label{def:intrinsic_totality}
  Totality for $\V$ is defined as~$v\in\T{\V}$ iff and only if every branch
  of~$v$ has even principal priority.
\end{defi}

Because of Corollary~\ref{cor:totality_values}, checking totality of a
recursive definition can thus take the following form:
\begin{enumerate}
  \item
    annotate the definition with priorities during type checking,

  \item
    check that the infinite unfolding of the recursive definition either

    \begin{enumerate}
      \item
        inspects a non-total infinite branch of the argument,

      \item
        or only constructs total infinite branches of the result.
    \end{enumerate}
\end{enumerate}
The patterns on the left side of clauses are the parts that ``inspect the
argument'' and the values on the right side of clauses are the parts that
``construct the result''.
A recursive definition satisfying this property is total. When applied to a
total value (which has no non-total branch), the result is necessarily total
(it contains only total branches).

Because we cannot really inspect the infinite unfolding of the definition, the
size-change principle will be used to give a computable approximation of the
above. Making this precise is the aim of an upcoming
paper~\cite{ph:SCP}.
%%%>>>1

%% vim600:set foldmarker=<<<,>>> foldmethod=marker fileencoding=ascii spelllang=en spell: %%

\section*{Concluding Remarks}

\subsubsection*{Operational Semantics}

We have voluntarily refrained from giving the operational semantics of the
language. The idea is that totality is a semantic property and the operational
semantics has to be compatible with the standard semantics of recursive
definitions.
The operational semantics must guarantee that evaluating a total function on a
total value is well defined, in particular that it should terminate. For
example, head reduction that stops on records guarantees that a total value
has a head normal form: it cannot contain~$\bot$ and cannot start with
infinitely many inductive constructors (their priority is odd). Evaluation
must reach a record (coinductive) at some point.

A real programming language could introduce two kinds of records: coinductive
ones and finite ones. The later could be evaluated during head reduction. Even
better, destructors themselves could be coinductive (like \tt{Tail} for
streams) or finite (like \tt{Head} for streams.)

In a similar vein, the language could have coinductive constructors to
deal with coinductive types like finite or infinite lists.\footnote{The
interaction between such coinductive constructors and dependent types is
however very subtle as they can break subject reduction!
\url{https://github.com/coq/coq/issues/6768}.} At the moment, the only way to
introduce this type is with
{\small\begin{alltt}
  data list_aux('a, 'b) where
     Nil : unit -> list_aux('a, 'b)
   | Cons : prod('a, 'b) -> list_aux('a, 'b)

  codata inf_list('a) where
     unfold : inf_list('a) -> list_aux('a, inf_list('a))
\end{alltt}}\noindent
Needless to say, using this quickly gets tiring.

\subsubsection*{Higher order types}
\label{sub:conclusion_HOT}

The implementation of \chariot does deal with some higher order
datatypes. With $T$-branching trees (coinductive) defined as
{\small\begin{alltt}
  codata tree('b, 'n) where
      child : tree('b, 'n) -> ('b -> tree('b, 'n))
\end{alltt}}\noindent
or (inductive)
{\small\begin{alltt}
  data tree('b, 'n) where
      root : unit -> tree('b, 'n)
    | fork : ('b -> tree('b, 'n)) -> tree('b, 'n)
\end{alltt}}\noindent
the corresponding \tt{map} function passes the totality test. The theory should
extend to account for this kind of datatypes.

%% vim600:set foldmarker=<<<,>>> foldmethod=marker fileencoding=ascii spelllang=en spell: %%

%%%<<<1 Bibliography
\newpage
\bibliographystyle{alphaurl}
\bibliography{chariot}

%%%>>>1

\end{document}